\documentclass[aip, amsmath, amssymb, superscriptaddress, reprint]{revtex4-1}
\usepackage{graphicx} 
\usepackage{dcolumn} 
\usepackage{bm} 
\usepackage{placeins} 
\usepackage{hyperref} 
\usepackage{hhline} 
\setcitestyle{super} 
\usepackage{color} 

\usepackage{comment} 

\usepackage[margin=1.0in]{geometry}
\usepackage{wrapfig} 

\begin{document}

\title{Crystalline shielding mitigates structural rearrangement and localizes memory in jammed systems under oscillatory shear}

\author{Erin G. Teich}
\affiliation{Department of Bioengineering, University of Pennsylvania, Philadelphia PA 19104 USA}
\author{K. Lawrence Galloway}
\affiliation{Department of Mechanical Engineering \& Applied Mechanics, University of Pennsylvania, Philadelphia PA 19104 USA}
\author{Paulo E. Arratia}
\affiliation{Department of Mechanical Engineering \& Applied Mechanics, University of Pennsylvania, Philadelphia PA 19104 USA}
\affiliation{Department of Chemical \& Biomolecular Engineering, University of Pennsylvania, Philadelphia PA 19104 USA}
\author{Danielle S. Bassett}
\affiliation{Department of Bioengineering, University of Pennsylvania, Philadelphia PA 19104 USA}
\affiliation{Department of Physics \& Astronomy, University of Pennsylvania, Philadelphia PA 19104 USA}
\affiliation{Department of Electrical \& Systems Engineering, University of Pennsylvania, Philadelphia PA 19104 USA}
\affiliation{Santa Fe Institute, Santa Fe NM 87501 USA}

\date{\today}

\begin{abstract}
The nature of yield in amorphous materials under stress has yet to be fully elucidated. 
In particular, understanding how microscopic rearrangement gives rise to macroscopic structural and rheological signatures in disordered systems is vital for the prediction and characterization of yield and for the study of how memory is stored in disordered materials.
Here, we investigate the evolution of local structural homogeneity on an individual particle level in amorphous jammed two-dimensional systems under oscillatory shear, and relate this evolution to rearrangement, memory, and macroscale rheological measurements. 
We identify a new structural metric, \emph{crystalline shielding}, that is predictive of rearrangement propensity and the structural volatility of individual particles under shear. 
We use this metric to identify localized regions of the system in which the material's memory of its preparation is preserved.
Our results contribute to a growing understanding of how local structure relates to dynamic response and memory in disordered systems.
\end{abstract}

\maketitle

\section{Introduction}
Amorphous, jammed systems \cite{Liu2010a} are abundant in nature and utilized often to process and produce materials \cite{Liu1998}.
The way in which these systems' disordered multi-scale structure \cite{Papadopoulos2018a} evolves under the application of stress \cite{Manning2011, Cubuk2017, Tordesillas2010, Herrera2011, Papadopoulos2016}, 
eventually resulting in catastrophic yielding \cite{Bonn2017}, 
is an area of active investigation with consequences for phenomena ranging from landslides and other forms of landscape evolution \cite{Ferdowsi2017, Jerolmack2019} to cellular unjamming during tumor metastasis \cite{Bi2016, Oswald2017}. The identification of local structural characteristics that are coupled to dynamical response under stress in disordered systems is of particular interest. Yet, associated efforts have been hampered by the lack of obvious structural order in such systems and the often subtle nature of the relevant dynamics.

Memory encoding in amorphous materials is an especially intriguing stress response with a highly non-trivial relationship to heterogeneous structure.
The evolution of a disordered system can depend on its preparation history, resulting in embedded memories of the past that can be read out by subsequent procedures \cite{Keim2019}. Material memory has been observed in a variety of forms ranging from the simple to the complex. For example, amorphous systems under oscillatory shear have recently been found to develop precisely cyclical particle trajectories \cite{Corte2008, Regev2013, Schreck2013, Keim2014, Royer2015, Lavrentovich2017} that can encode single or multiple memories of the strain amplitude at which the material was prepared  \cite{Keim2011, Paulsen2014, Fiocco2014, Adhikari2018, Mukherji2019, Keim2020}. Materials may also simply remember the direction in which they were last deformed and express that memory via directional asymmetry in their response to subsequent stress, as in the case of the well-known Bauschinger effect observed in metals \cite{Dieter1961, Sowerby1979} and amorphous materials \cite{Falk1998, Karmakar2010, Bhattacharjee2015}. Even in such a seemingly simple case, however, important open questions remain regarding the microstructural origin of response asymmetry \cite{Karmakar2010}. 
To understand the more complicated forms of memory in materials, it is crucial that the relationship between local structure and the simple memory encoded by shear response anisotropy in heterogeneous materials be fully elucidated.

Here, we provide a conceptual link between local structure and memory as encoded in the response of a two-dimensional dense colloidal system under oscillatory shear.
Experiments are performed using a custom-made interfacial stress rheometer in which a dense (jammed) particle monolayer is sheared using a magnetic needle \cite{Keim2015}. This method provides sufficient spatial and temporal resolution to probe structural rearrangement probability on the scale of individual particles, while simultaneously measuring bulk rheological properties.

We show that these jammed and disordered systems have memory of their preparation, exhibited via an asymmetry in local deformation with respect to shear direction below yielding. We use generic measures of structural homogeneity on an individual particle level to track the sample microstructure over time, and find that global crystallinity is also asymmetric with respect to shear direction, and thus encodes preparation memory. 
This memory is increasingly ``erased" as strain amplitude increases beyond yielding.
We next demonstrate that, on an individual particle level, correlations in crystallinity over time are reliable indicators of particle rearrangement. 
We stratify crystalline particles into subgroups according to their interiority within crystal grains, and find that the propensity for rearrangement occurs in a hierarchy according to this crystalline shielding metric, with particles most interior within grains rearranging least. 
Our results show that the likelihood of particle rearrangement depends on a continuum of interiority within crystal grains rather than a binary classification of grain boundary vs. interior as has been found previously \cite{Gokhale2012, Keim2014, Keim2015, Buttinoni2017, Sharp2018}. Finally, we show that rearrangement asymmetry with respect to shear direction also occurs in a hierarchy according to crystalline shielding, with asymmetry being highest for particles most interior within grains. 
Thus, we conclude that sample preparation memory is spatially localized to the interior of crystal grains.

\section{Methods}
\subsection{Experiments}
Experiments are performed using a custom-made interfacial stress rheometer (ISR) to controllably impose shear deformation on two-dimensional jammed colloidal suspensions. The ISR apparatus allows for the tracking of single particles (and hence microstructure characterization), while simultaneously measuring the suspension bulk rheological properties (e.g. viscous and elastic modulii). We will briefly describe the ISR employed in our experiments, and further details can be found in Refs. \citenum{Keim2014, Keim2015}. The ISR is composed of a ferromagnetic needle trapped at a decane/water interface by capillary forces, between two vertical glass walls. 
These walls pin the interface to maintain a flat shearing channel that can be simultaneously imaged with a microscope. Particles are adsorbed at this interface, creating a two-dimensional jammed colloidal suspension. Particle positions are identified and linked to form trajectories using the open-source particle-tracking software \emph{trackpy} \cite{trackpy, Crocker1996}. The positions of approximately 40,000 particles are tracked during shearing.

To obtain an interface's rheological information, the needle is driven axially by a known, imposed, magnetic force generated by two Helmholtz coils. The displacement of the needle is measured using a microscope. 
A monolayer's shear storage ($G'$) and loss ($G''$) moduli are calculated from the imposed force and observed displacement \cite{Shahin1986, Brooks1999a, Reynaert2008}. Experiments access shear moduli over a range of strain amplitudes $\gamma_0$, with $0.005 < \gamma_0 < 0.16$, at a fixed frequency of 0.1 Hz. Prior to each experiment, the monolayer is prepared \emph{via} six cycles of shearing at a large strain amplitude ($\gamma_0 \sim 0.5$). Shearing is then halted, and resumed at smaller strain amplitudes for experimental data collection.

All particles are sulfate latex (invitrogen) and experience dipole-dipole repulsion due to charge groups on the surface of the particles \cite{Park2010}. All packings have high enough area fraction $\phi$ to be fully jammed without shear. Data from two main experimental systems are analyzed here: a bi-disperse system (equal parts 4.1$\mu m$ and 5.6$\mu m$ diameters, $\phi \approx 43\% $) is analyzed in the main \emph{Results} section, and a mono-disperse system (5.6$\mu m$ diameter, $\phi \approx 32\% $) is analyzed in the \emph{Discussion} and \emph{Supplementary Information} sections. (We note that these experimental systems were examined in Ref.~\citenum{Keim2015} with different analytical techniques to test distinct hypotheses.)
A snapshot of a bi-disperse two-dimensional jammed amorphous colloidal system is shown in Fig. \ref{fig:methods}A. Oscillatory rheology measured for these systems is shown in Fig. \ref{fig:methods}B.
In both systems, the rheological yield strain amplitude is approximately $\gamma_0 \sim 0.03$.

\subsection{Structural analysis}
\begin{figure*}
\centering
\includegraphics[width=\textwidth]{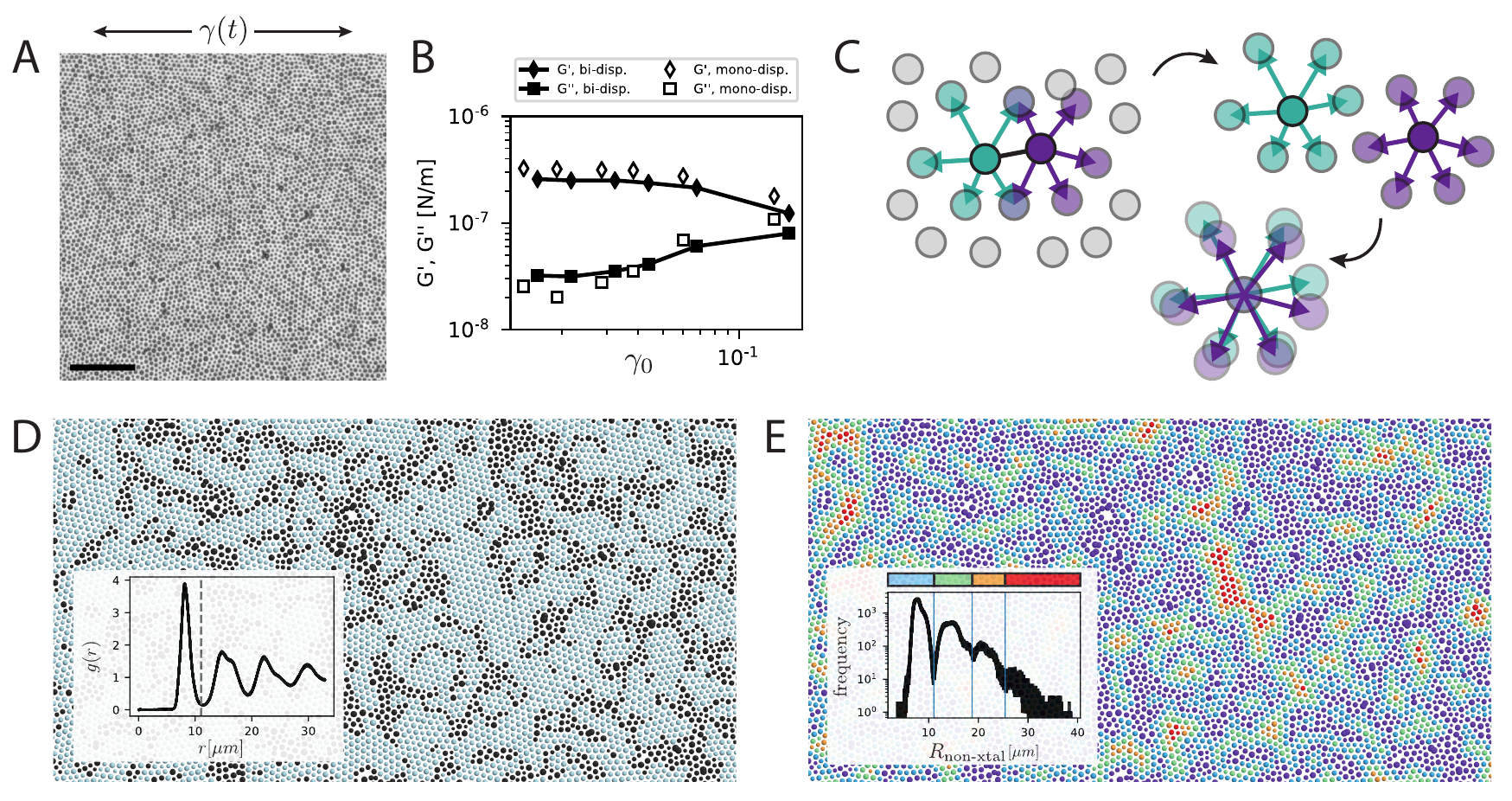}
\caption{\textbf{The system and structural characterization methods used in this work.} 
(A) Snapshot of a two-dimensional jammed amorphous bi-disperse colloidal system. The scale bar is 100 $\mu m$.
An arrow shows the direction of oscillating strain $\gamma (t)$.
(B) Oscillatory rheology also reported in Ref. \citenum{Keim2015}. Connected symbols are measurements for the bi-disperse systems considered in the \emph{Results} section, and unconnected symbols are measurements for the mono-disperse systems mentioned in the \emph{Discussion} section and presented fully in the \emph{Supplementary Information}.
(C) Schematic illustrating the environment matching method used to determine crystallinity. 
We identify neighbors, extract their environments, and then compare those environments to determine local structural homogeneity.
(D) Visual rendering of a portion of a sample bi-disperse experiment at $\gamma_0 = 0.068$. Here, particles in crystalline grains of population greater than 1 are colored light blue. 
The inset is the radial distribution function collected over all particles during one cycle of the experiment, with the nearest-neighbor distance $r_{cut}$ marked in gray. 
(E) Visual rendering of the identical system snapshot with particles colored according to crystalline shielding $R_{\text{non-xtal}}$. 
Disordered particles are colored purple.
The inset is a histogram of $R_{\text{non-xtal}}$ collected over all particles during 6 cycles of the experiment.
}
\label{fig:methods}
\end{figure*}

To investigate local structure in these amorphous colloidal systems, we use an environment matching method to characterize continuous structural homogeneity and discrete crystallinity.
Software can be found in the open-source analysis toolkit \emph{freud} \cite{Ramasubramani2020}. 
Our approach, schematically illustrated in Fig. \ref{fig:methods}C, characterizes whether particle environments are sufficiently \emph{similar} to their neighbors' environments, regardless of the structure of the environment itself.

We define particle $i$'s environment (shown as green vectors in Fig. \ref{fig:methods}C) as the set of vectors $\{ \bm{r}_{im} \}$, where $\bm{r}_{im}$ points from the center of particle $i$ to the center of particle $m$, and $m$ is an index over $i$'s $M_i$ nearest neighbors. 
We then inspect all environments of $i$'s neighbors. Let one such neighbor be labeled $j$. Particle $j$'s environment (shown as purple vectors in Fig. \ref{fig:methods}C) is defined as the set of vectors $\{ \bm{r}_{jm'} \}$, where $\bm{r}_{jm'}$ points from the center of particle $j$ to the center of particle $m'$ and $m'$ loops over $j$'s $M_j$ nearest neighbors. 
We then compare the environments of particle $i$ and particle $j$ by attempting to match these sets of vectors. Particle $j$'s environment ``matches" particle $i$'s environment if we can find a one-to-one mapping such that $\vert \bm{r}_{im} - \bm{r}_{jm'} \vert < t$ for every mapping pair $(m,m')$ for some threshold $t$.
Nearest neighbors of each particle are defined as those within a radial distance $r_{cut} = 11.04 \mu m$ for the bi-disperse systems and $r_{cut} = 11.34 \mu m$ for the mono-disperse systems, determined approximately by the minimum after the first peak of the radial distribution function $g(r)$ calculated over all particles in each system. Fig. \ref{fig:methods}D (inset) shows the radial distribution function $g(r)$ of a sample bi-disperse experiment at $\gamma_0 = 0.068$, collected over one shear cycle. Environments of particles $i$ and $j$ are only compared if $M_i = M_j$, and thus a one-to-one mapping is possible; otherwise, particles $i$ and $j$ are automatically deemed non-matching. The threshold $t = 0.2 r_{cut}$ was chosen for all systems, because 0.2 or 0.3 times the approximate nearest-neighbor distance ($r_{cut}$) has proven appropriate -- neither too stringent nor too lenient -- in other contexts \cite{Cadotte2016,Teich2019}.
Supplementary Fig. S2 explores the impact of threshold choice on crystallinity characterization as we explain next.

If the environments of two neighboring particles match, then they are designated members of the same crystal grain. 
Crystallinity is defined in this paper as the fraction of particles in crystal grains of size larger than 1, and particles are defined as crystalline if they are members of a crystal grain of size larger than 1.
Fig. \ref{fig:methods}D shows a snapshot of a sample bi-disperse experiment at $\gamma_0 = 0.068$ with all crystalline particles identified.
Particles are drawn with radii equal to twice the measured image radii of gyration for ease of visualization.
Crystalline structure in this system is hexagonal in nature, as reported in Ref. \citenum{Keim2015} and also as evidenced by histograms of the bond-orientational order parameter $\vert \psi_6 \vert$, collected separately for crystalline particles and non-crystalline particles over one cycle of two example systems (Supplementary Fig. S1).
The complex number $\psi_6 (i) = \frac{1}{N_i} \sum_{j=1}^{N_i} e^{6i\phi_{ij}}$, where $N_i$ is the number of nearest neighbors of particle $i$ and $\phi_{ij}$ is the angle between $\bm{r}_{ij}$ and the vector $(1,0)$, measures the six-fold orientational symmetry of particle $i$'s environment.
The distribution of $\vert \psi_6 \vert$ values for crystalline particles peaks near 1 in both experiments shown in Fig. S1, implying strong hexagonal order, while $\vert \psi_6 \vert$ for disordered particles is distributed evenly across all values between 0 and 1.
The choice of matching threshold $t$ influences the hexagonal quality of the identified crystal grains as shown in Supplementary Fig. S2A. We find that setting the threshold below $t = 0.2 r_{cut}$ results in many particles being deemed disordered that nevertheless have high hexagonal ordering, whereas setting the threshold above $t = 0.2 r_{cut}$ results in many particles being deemed crystalline that have low hexagonal ordering.
Thus, $t = 0.2 r_{cut}$ is a reasonable compromise that produces a well-defined bipartition of crystalline and disordered particles.
Furthermore, we find that the choice of threshold, within reason, does not change our results pertaining to crystallinity reported in Section \ref{section:asymm} (see Figs. S2B, S2C).

We quantify the shielding, or interiority of a crystalline particle within a grain, by $R_{\text{non-xtal}}$, the distance of that particle to the nearest non-crystalline particle. 
Shielding is higher as $R_{\text{non-xtal}}$ increases.
Example distributions of $R_{\text{non-xtal}}$ for all particles in 6 consecutive non-transient cycles of a sample bi-disperse experiment at $\gamma_0 = 0.068$ (Fig. \ref{fig:methods}E inset) show clear peaks and valleys that inform the way in which we bin $R_{\text{non-xtal}}$ into four shielding levels, $\{ R_i \}$. 
The minima of $R_{\text{non-xtal}}$ are approximately the first three minima of $g(r)$, shown inset in Fig. \ref{fig:methods}D, as one would expect. 
These minima are not influenced by strain amplitude, as shown in Supplementary Fig. S7, which displays distributions of $R_{\text{non-xtal}}$ for all particles over several cycles in experiments below and above yield.
A rendering of the system with particles colored according to crystalline shielding illustrates the concept (Fig. \ref{fig:methods}E).
Disordered particles, whose distance to the nearest non-crystalline particle is formally zero, are also shown and colored purple.

Due to occasional imaging or tracking errors, or to particles moving out of the imaging field of view, some particles are not preserved over the course of entire experiments. 
The fraction of preserved particles ranges from $\sim 0.97$ (for the smallest strain amplitude) to $\sim 0.87$ (for the largest strain amplitude) in the bi-disperse system, and from $\sim 0.61$ to $\sim 0.85$ (not correlated with strain amplitude) in the mono-disperse system.
To eliminate any spurious effects due to the non-preserved particles, we typically calculate structural signatures of all particles in every snapshot, but only show those signatures of particles that are tracked and preserved over the entire experiment. 
Unless otherwise stated, the following results will always be for the preserved particles in each experiment.
 
\section{Results}
\subsection{Crystallinity oscillates with shear and reveals material preparation memory} \label{section:asymm}
We first investigate whether our systems retain memory of their preparation history, by studying the symmetry of their deformation response with respect to shear direction. 
We measure local deformation according to how much the average nearest neighbor shell stretches, and find that deformation is asymmetric with respect to shear direction at low strain amplitude. Local deformation displays increasing symmetry as strain amplitude increases. 

\begin{figure*}
\centering
\includegraphics[width=0.9\textwidth]{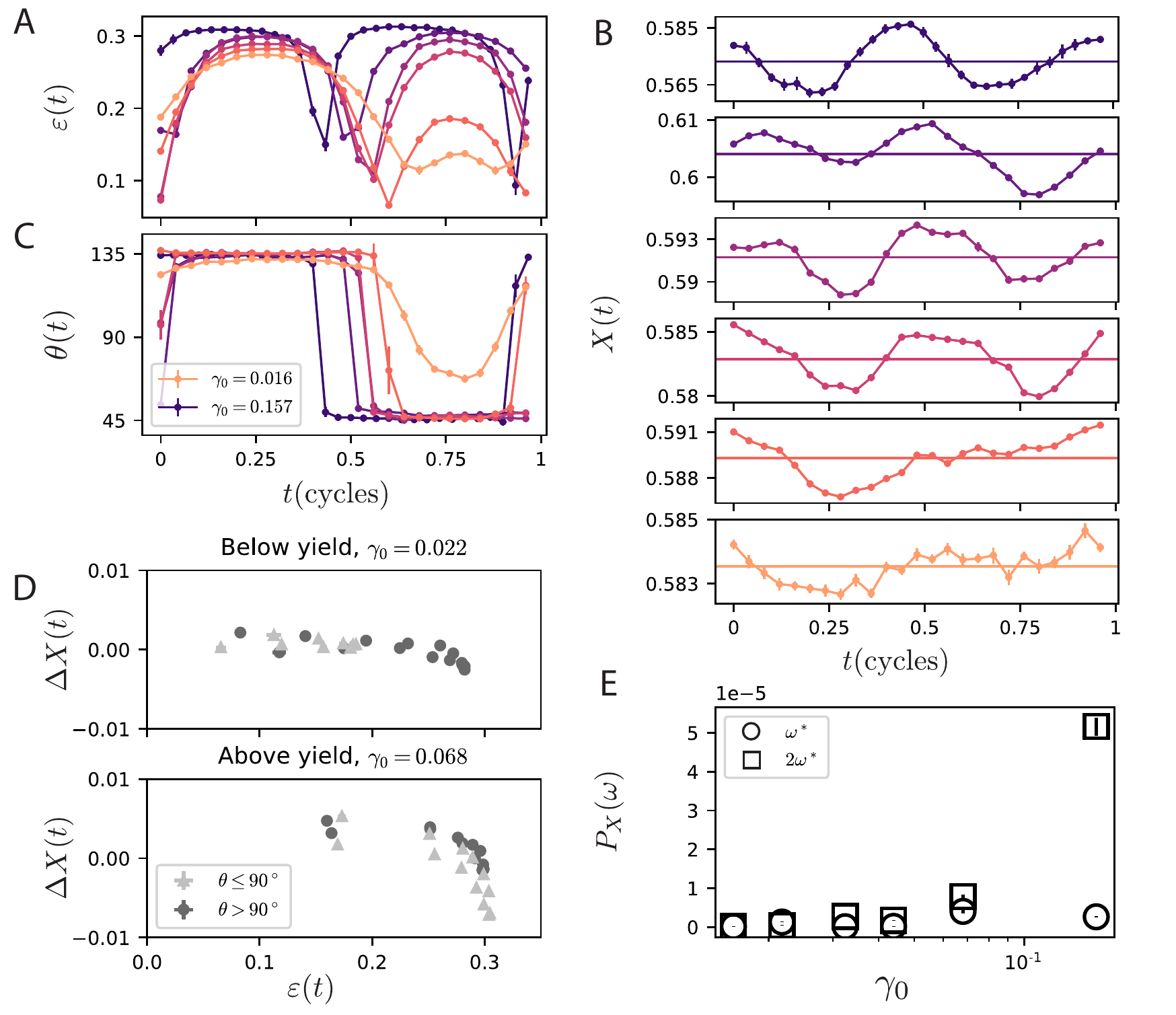}
\caption{\textbf{Global signatures of crystallinity and local neighborhood deformation as a function of strain amplitude $\gamma_0$ show asymmetry with respect to shear direction.}
(A,C) Stroboscopic averages of local neighborhood deformation. The quantity $\varepsilon$ is the eccentricity of the local neighborhood ellipse, and $\theta$ is its orientation. 
Error bars represent the standard error of the mean value at each time point.
Increasingly dark colors correspond to increasing values of $\gamma_0$.
Values of $\gamma_0$ for experiments at the lowest and highest strain amplitudes studied are shown. 
(B) Crystallinity in all systems, stroboscopically averaged. 
Error bars again represent the standard error of the mean value at each time point.
Increasingly dark colors correspond to increasing values of $\gamma_0$. 
Horizontal lines indicate the mean of each crystallinity signature. 
(D) Deviations in the crystallinity from its mean, $\Delta X(t)$, plotted against local neighborhood deformation $\varepsilon (t)$ for example systems below and above yield.
Shown are stroboscopically averaged quantities and corresponding error bars identical to those in panels A and B.
Light gray triangles mark all frames during the second shear half-cycle, for which $\theta(t) \leq 90^\circ$, and dark gray circles mark all frames during the first shear half-cycle, for which $\theta(t) > 90^\circ$.
(E) Two power spectral densities $P_X(\omega)$ of non-transient crystallinity signatures as a function of $\gamma_0$, for two distinct frequencies. The circles correspond to frequency $\omega^*$, which is the frequency of the needle oscillation; the squares correspond to frequency $2\omega^*$, which is the second harmonic of the needle oscillation. 
Error bar estimation is described in the text.
}
\label{fig:xtal}
\end{figure*}

Fig. \ref{fig:xtal}A shows the eccentricity of the ellipse fit to the average nearest neighbor shell over multiple shear cycles for experiments at all strain amplitudes, and Fig. \ref{fig:xtal}C shows ellipse orientation, defined as the angle from the positive x-axis to its major axis.
Eccentricity and orientation signatures are calculated for each experiment from an ellipse fit to the average nearest neighbor shell of all particles (even if they are not preserved over the entire experiment), defined as the boundary demarcated by the first peak of the two-dimensional histogram of nearest neighbors accumulated over every particle.
We show stroboscopic averages of each signal over the non-transient portion of each system trajectory, conservatively defined as the set of cycles well after the global crystallinity for each system, defined below, has reached steady-state oscillation.
Signals for full trajectories are shown in Supplementary Fig. S3. 

At high strain amplitude, well above yield, eccentricity reaches approximately equal heights during the first and second halves of each cycle, and thus local deformation is approximately symmetric with respect to shear direction.
As strain amplitude decreases, however, eccentricity grows smaller (and thus local deformation is smaller) during the back half of each cycle, when $\theta \sim 45^\circ$. 
Asymmetry in deformation between the first and second shear half-cycles implies an anisotropy in the system that encodes the system's history, and this memory is ``erased" with increasing strain amplitude.   

We next consider whether this asymmetrical response is exhibited by measures of structure in our systems.
We calculate global crystallinity for each system over time and find that it also displays an asymmetry with respect to shear direction at low strain amplitude, thus seeming to also indicate material anisotropy and preparation memory.
Global crystallinity at time $t$, $X(t)$, is the fraction of particles in the system that are crystalline, as defined in the \emph{Methods}.
We note that we could instead have chosen to analyze a more continuous per-particle measure of structural homogeneity, $\Delta_i$: the root-mean-squared deviation between the environment of particle $i$ and that of its neighbor, averaged over all of its neighbors. Another global measure of structural homogeneity is then $\overline{\Delta}(t)$, where the average is taken over all particles in the system at each time. 
Calculation of $\Delta$ is explained in more depth in the \emph{Supplementary Information}, and results using this alternative definition, shown in Supplementary Fig. S5, are very similar to those presented here.

Fig. \ref{fig:xtal}B shows that global crystallinity $X(t)$ oscillates in time with shear, in agreement with other studies \cite{Zhang2010} that observed similar structural oscillations. 
The amplitude of the crystallinity oscillation increases with strain amplitude.
We show stroboscopic averages of each signal over non-transient trajectories identical to those used to calculate Fig. \ref{fig:xtal}A, C. 
Full signals, showing initial transient behavior, are shown in Supplementary Fig. S3.
At low strain amplitude, there is an asymmetry in the crystallinity signal with respect to shear direction, and this asymmetry is erased as strain amplitude increases. 
This asymmetry erasure is evident in the power spectra of the signals via the periodogram estimate (Fig. \ref{fig:xtal}E): we find that the power spectral density associated with twice the frequency of the needle oscillation, $P_X (2\omega^*)$, increases with strain amplitude, whereas $P_X (\omega^*)$ remains relatively stable.
Each power spectral density is the mean of a set of $P_X(\omega)$ values calculated over consecutive 2 cycle windows of the relevant non-transient trajectory shown in Supplementary Fig. S4, and error bars represent the standard deviation of the mean.
Full power spectral density distributions over all $\omega$ are also shown in Supplementary Fig. S4. 

In general, global crystallinity decreases as deformation increases. 
This behavior can be seen in plots of deviation in the crystallinity from its mean, $\Delta X(t) \equiv X(t) - \langle X \rangle_t$, against neighborhood ellipse eccentricity for systems below and above yield (Fig. \ref{fig:xtal}D).
Plots for all experiments are shown in Supplementary Fig. S6.
As eccentricity increases, crystallinity dips, and this dip is more pronounced above yield.
Below yield, crystallinity during the second half-cycle (for which $\theta(t) \leq 90^\circ$) remains distributed close to $\Delta X(t) = 0$, due to asymmetry with respect to shear direction. 
In the experiment we show above yield, crystallinity during the second half-cycle dips even lower than crystallinity during the first half-cycle (for which $\theta(t) > 90^\circ$).

\subsection{Correlations in crystallinity over time indicate particle rearrangement under shear}
A closer investigation of the structure of individual particles over time reveals that correlations in crystallinity are reliable indicators of individual particle rearrangement. To show this, we quantify crystallinity correlation via $p(s,t \vert s, t_0)$, the conditional probability that a particle is in structure $s$ at time $t$ given that it was in the same structure $s$ at time $t_0$. In our analysis, either $s=x$, representing crystalline structure, or $s=d$, representing non-crystalline or disordered structure. We compare crystallinity correlation to a metric that measures particle rearrangement: $D^2_{min}(t_0,t)$ \cite{Falk1998}, the mean squared deviation of the displacements of a particle and its neighbors from the best-fit affine deformation of those displacements \cite{philatracks}. High values of $D^2_{min}(t_0,t)$ correspond to non-affine deformations between times $t_0$ and $t$, which manifest as particle rearrangements. We find that $p(s,t \vert s, t_0)$ and $D^2_{min}(t_0,t)$ are inversely related: when $p(s,t \vert s, t_0)$ is lower, $D^2_{min}(t_0,t)$ is higher, and \emph{vice versa}. This relationship implies that structural auto-correlation captures particle rearrangement dynamics.

To gain more insight into the influence of crystalline structure on rearrangement dynamics, we partition particles into groups according to their crystalline shielding level at time $t_0$, as explained in the \emph{Methods}, when calculating $p(s,t \vert s, t_0)$ and $D^2_{min}(t_0,t)$.
Supplementary Figs. S7 and S8 show fractions of particles at each crystalline shielding level as a function of time for all strain amplitudes; we observe that each signature oscillates distinctly in time, showcasing the evolution of crystal grain morphology during the shear cycle.

\begin{figure*}
\centering
\includegraphics[width=0.8\textwidth]{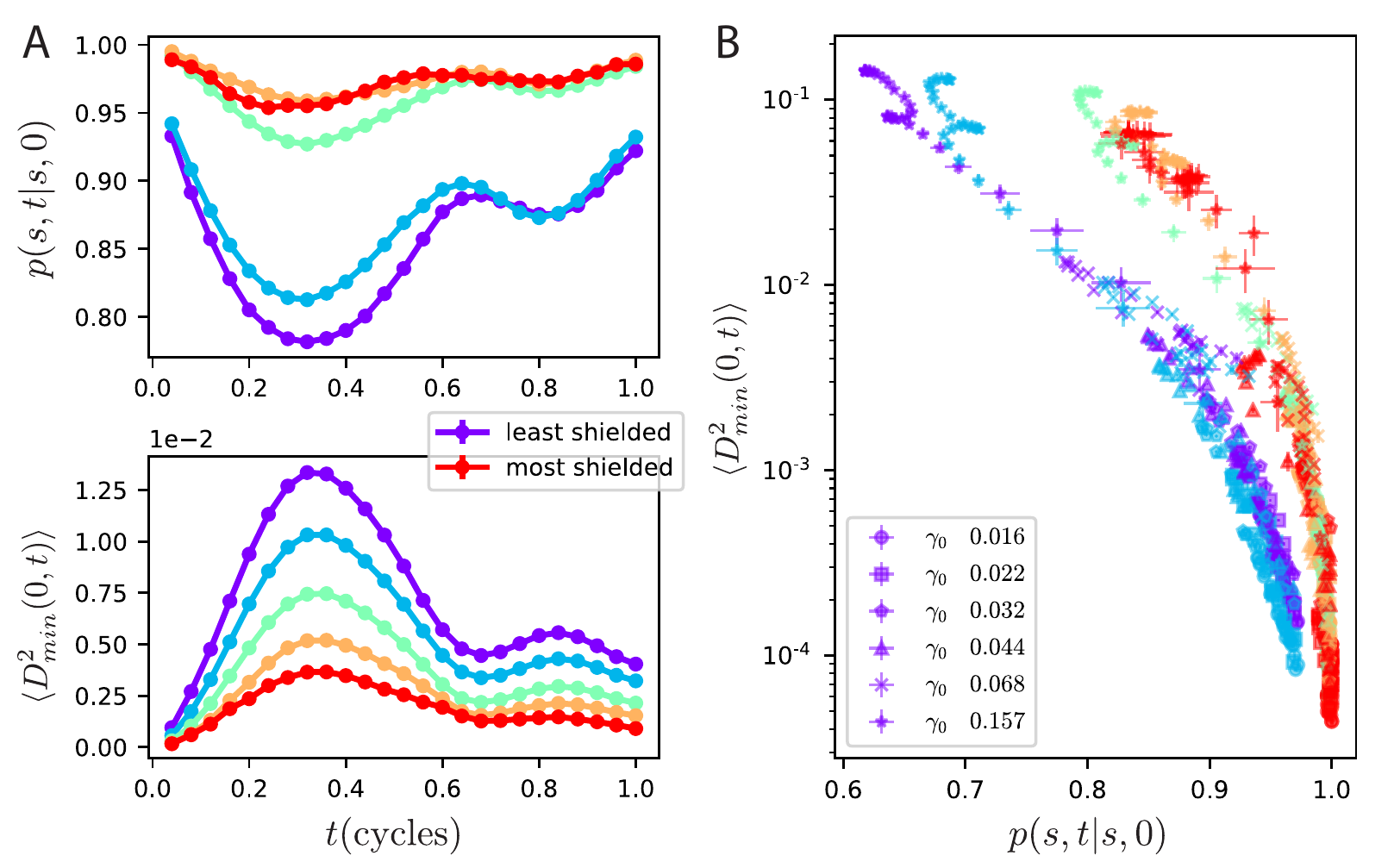}
\caption{
\textbf{Correlations in crystallinity indicate individual particle rearrangement.}
(A) Rearrangement measurements $p(s,t \vert s, 0)$ (top) and $\langle D^2_{min}(0,t) \rangle$ (bottom) for a sample experiment at $\gamma_0 = 0.068$.
Signals are shown as a function of $t$ over one shear cycle, and $t_0 = 0$ marks the beginning of the cycle.
Results are shown as stroboscopic averages, and error bars represent the standard error of the mean value at each time point.
Each signal represents a shielding level according to the color scheme detailed in Fig. \ref{fig:methods}, and is calculated over particles in the appropriate shielding level at $t_0$.
Colors of the least shielded (disordered) and most shielded layers are shown for reference.
(B) Stroboscopically averaged $\langle D^2_{min}(0,t) \rangle$ as a function of stroboscopically averaged $p(s,t \vert s, 0)$, for all strain amplitudes and all shielding layers. 
Error bars are standard errors of the mean in both dimensions.
}
\label{fig:d2min}
\end{figure*}

We find that rearrangements occur in a hierarchy according to shielding level, with more shielded particles prone to less rearrangement at all points of the shear cycle. This result indicates that degree of interiority within crystal grains has a significant impact on rearrangement dynamics. As an example, we consider a sample experiment at $\gamma_0 = 0.068$ (above yield) in Fig. \ref{fig:d2min}A. The top panel of Fig. \ref{fig:d2min}A shows the quantity  $p(s,t \vert s, 0)$ for each shielding layer as a function of $t$ over one shear cycle, with $t_0 = 0$ marking the beginning of the cycle. The bottom panel shows the quantity $\langle D^2_{min}(0,t) \rangle$ for the same experiment and identical values of $t_0 = 0$ and $t$, where the average is taken over all particles in each shielding level at $t_0 = 0$. Both panels show stroboscopic averages of each signal over non-transient trajectories. While particles in all shielding layers show rearrangement within the shear cycle according to both $p(s,t \vert s, 0)$ and $\langle D^2_{min}(0,t) \rangle$, we find that more interior shielding layers generally show higher values of $p(s,t \vert s, 0)$ and lower values of $\langle D^2_{min}(0,t) \rangle$, indicating less rearrangement.

Additionally, in all cases, $p(s,t \vert s, 0)$ reaches a global minimum and $\langle D^2_{min}(0,t) \rangle$ reaches a global maximum around $t=0.25$ cycles (the time of the first strain extremum), while these signals reach local minima and local maxima, respectively, around $t=0.75$ cycles (the time of the second strain extremum). 
Rearrangement is more evident during the first shear half-cycle; this asymmetry is further evidence of a difference in material response according to shear direction, and hints that the system remembers its history.

We compile results for all strain amplitudes in Fig. \ref{fig:d2min}C, which clearly evidences the inverse relationship between $p(s,t \vert s, 0)$ and $\langle D^2_{min}(0,t) \rangle$ for all shielding levels.
As strain amplitude increases, the minima in $p(s,t \vert s, 0)$ decrease, and the maxima in $\langle D^2_{min}(0,t) \rangle$ increase, indicating increased rearrangement with strain amplitude.
Curiously, shielding levels show a roughly log-linear relationship between $\log \langle D^2_{min}(0,t) \rangle$ and $\left( 1 -  p(s,t \vert s, 0) \right)$ across most strain amplitudes, with disordered particles and particles at grain boundaries showing a smaller slope between these quantities than the more interior crystalline particles.
We may conclude in general that correlations in crystallinity over time, captured by $p(s,t \vert s, t_0)$, indicate rearrangement of individual particles, and that rearrangement occurs in a hierarchy according to interiority within crystal grains.

Finally, we note that these results seem quite robust; alternate analysis of structural rearrangement using time correlations in the continuous structural homogeneity parameter $\Delta_i$, introduced in Section \ref{section:asymm}, leads to similar conclusions to those presented above (see Supplementary Fig. S9). 

\subsection{Asymmetry in crystallinity correlation with respect to shear direction is localized within crystal grains}
Next, we demonstrate that particles more interior within crystal grains, or more shielded, have a rearrangement propensity that is more asymmetric with respect to shear direction, indicating that memory of material history is localized within crystal grains. To do so, we quantify the difference in rearrangement propensity with respect to shear direction via the correlation $p(s, t \vert s, t-0.5)$, the conditional probability that a particle is in structure $s$ (either crystalline or disordered) at time $t$ given that it was in the same structure $s$ half a cycle earlier. 
The quantity $p(s, t \vert s, t-0.5)$ thus measures correlation between equivalent time points in each half-cycle, differing only in the direction of shear, since the applied shear is sinusoidal.
As in the previous section, at each time $t$, we group particles according to their crystalline shielding level at time $t_0 = t-0.5$, and calculate $p(s, t \vert s, t-0.5)$ over each particle subgroup.
(Note that this correlation, and the correlations presented in the previous section, are in fact one-dimensional cuts through a full two-dimensional probability distribution $p(s, t \vert s, t_0)$. Full two-dimensional distributions for non-transient portions of trajectories at all strain amplitudes are shown in Supplementary Figs. S10 and S12.)

Fig. \ref{fig:corr}A shows $p(s, t \vert s, t-0.5)$ for all shielding layers for experiments below ($\gamma_0 = 0.022$) and above ($\gamma_0 = 0.068$) yield.
Signals are stroboscopic averages over non-transient system trajectories shown in Supplementary Fig. S12.
The two minima in each signal show that particles in all shielding levels are least structurally auto-correlated when both $t$ and $t-0.5$ are times of strain extremum, at 0.25 and 0.75 cycles.
This bimodal nature stems from the fact that particles are most dynamically responsive to shear, rearranging most, at those times of strain extremum.
However, for the more shielded layers in the experiment below yield, there is a striking asymmetry between the first and second halves of the signal.
The correlation $p(s, t \vert s, t-0.5)$ reaches a shallower minimum in the second half-cycle, around $t=0.75$ cycles, than it does in the first half-cycle, around $t=0.25$ cycles.
This asymmetry implies that shielded particles are less responsive to shear during the second half-cycle than they are during the first half-cycle.
The asymmetry is not as prominent for the least shielded layers, and diminishes in the experiment above yield.

We quantify this asymmetry via $P_p (2\omega^*)/P_p (\omega^*)$, the ratio of power spectra of each $p(s, t \vert s, t-0.5)$ signal at $2\omega^*$ and $\omega^*$, where $\omega^*$ is the frequency of the shear cycle. Fig. \ref{fig:corr}B shows this ratio for each shielding layer at all strain amplitudes, and provides further evidence that more shielded crystalline particles are more asymmetric with respect to direction in their response to shear.
Power spectra are calculated via the periodogram estimate and each value $P_p (\omega)$ is the mean of a set of such values calculated over consecutive two-cycle windows of the full non-transient $p(s, t \vert s, t-0.5)$ signal shown in Supplementary Fig. S12.
Power spectral densities for each frequency at all strain amplitudes are reported separately in Supplementary Fig. S11.
The ratio $P_p (2\omega^*)/P_p (\omega^*)$ is highest for the least shielded particles, and lowest for the most shielded particles, at all strain amplitudes. This finding implies that the structural rearrangement of the least shielded particles is most symmetric with respect to shear direction, and the structural rearrangement of the most shielded particles is least symmetric with respect to shear direction.
In fact, our results suggest a hierarchy of asymmetry in structural response according to crystalline shielding layer.
Thus, response anisotropy, signifying the material's memory of its history, is spatially localized within crystal grains.

\begin{figure*}
\centering
\includegraphics[width=0.8\textwidth]{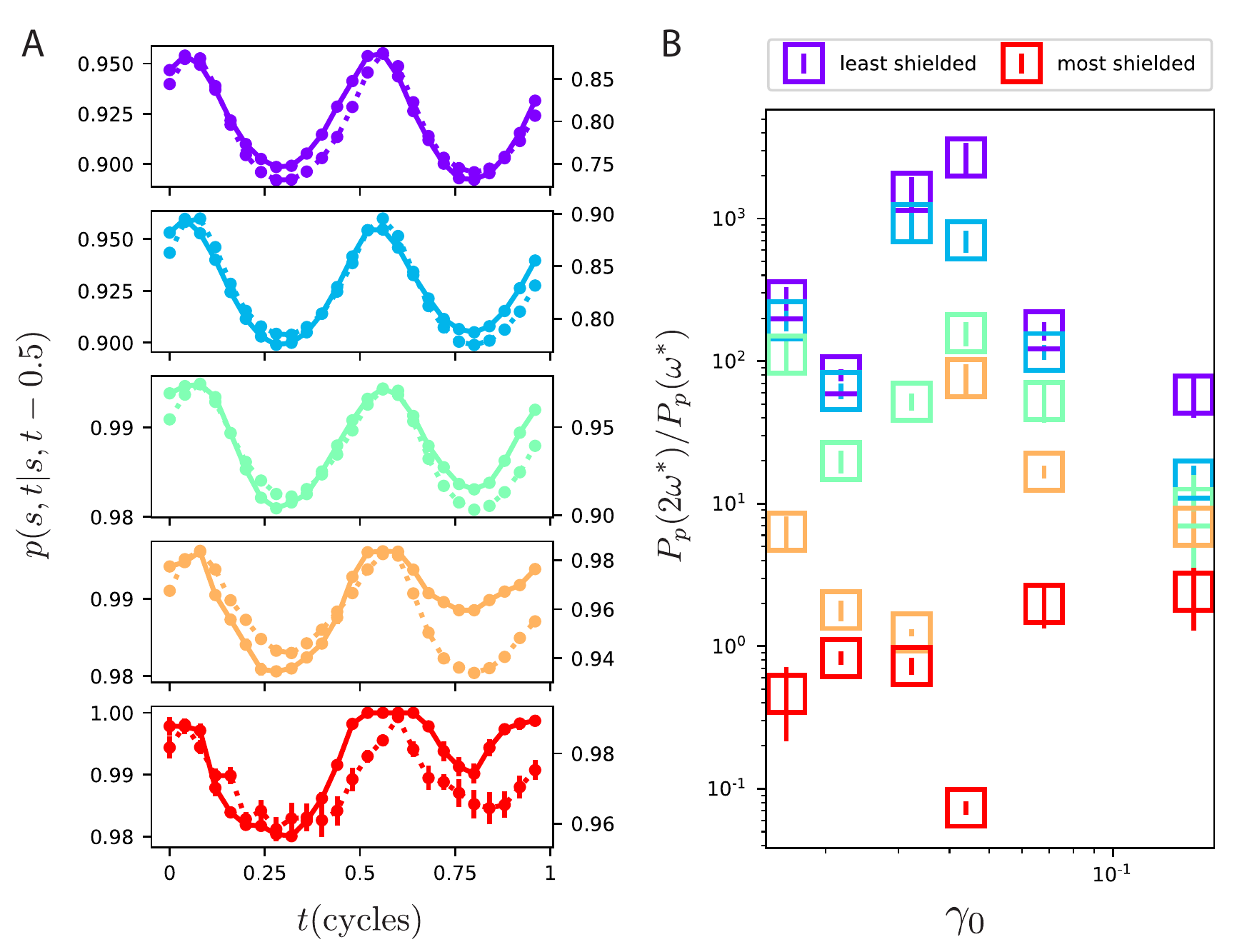}
\caption{
\textbf{More shielded crystalline particles are more asymmetric in structural response with respect to shear direction.}
(A) Structural responses $p(s, t \vert s, t-0.5)$ for particles at all shielding levels for experiments below ($\gamma_0 = 0.022$) and above ($\gamma_0 = 0.068$) yield, as solid and dotted lines, respectively.
Stroboscopic averages are shown for clarity, and error bars denote the standard error of the mean value at each time point.
(B) The ratio $P_p (2\omega^*)/P_p (\omega^*)$ for all shielding layers as a function of $\gamma_0$. 
Error bars are calculated via Taylor series propagation of the standard errors of the mean of each $P_p (\omega)$ quantity in the ratio.
In both panels, signals are colored by shielding level according to the color scheme detailed in Fig. \ref{fig:methods}. 
Colors of the least shielded (disordered) and most shielded layers are shown for reference.
} 
\label{fig:corr}
\end{figure*}

\section{Discussion}
We have demonstrated that bi-disperse amorphous, jammed systems under oscillatory shear show an asymmetry with respect to shear direction in both local deformation and structural homogeneity. This asymmetric response is only erased at high strain amplitudes above yield. Per-particle auto-correlations in structural homogeneity are also asymmetric with respect to shear direction, and auto-correlations are especially asymmetric for particles that are most interior, \emph{i.e.} shielded, within crystal grains. We believe that observed asymmetries are indicative of memory of the system's history or preparation, and our findings imply that this simple form of memory is spatially localized within crystal grains.

\subsection{Structural reversibility is destroyed only at strain amplitudes well above yield}  \label{section:irrev}
\begin{figure*}
\centering
\includegraphics[width=0.6\textwidth]{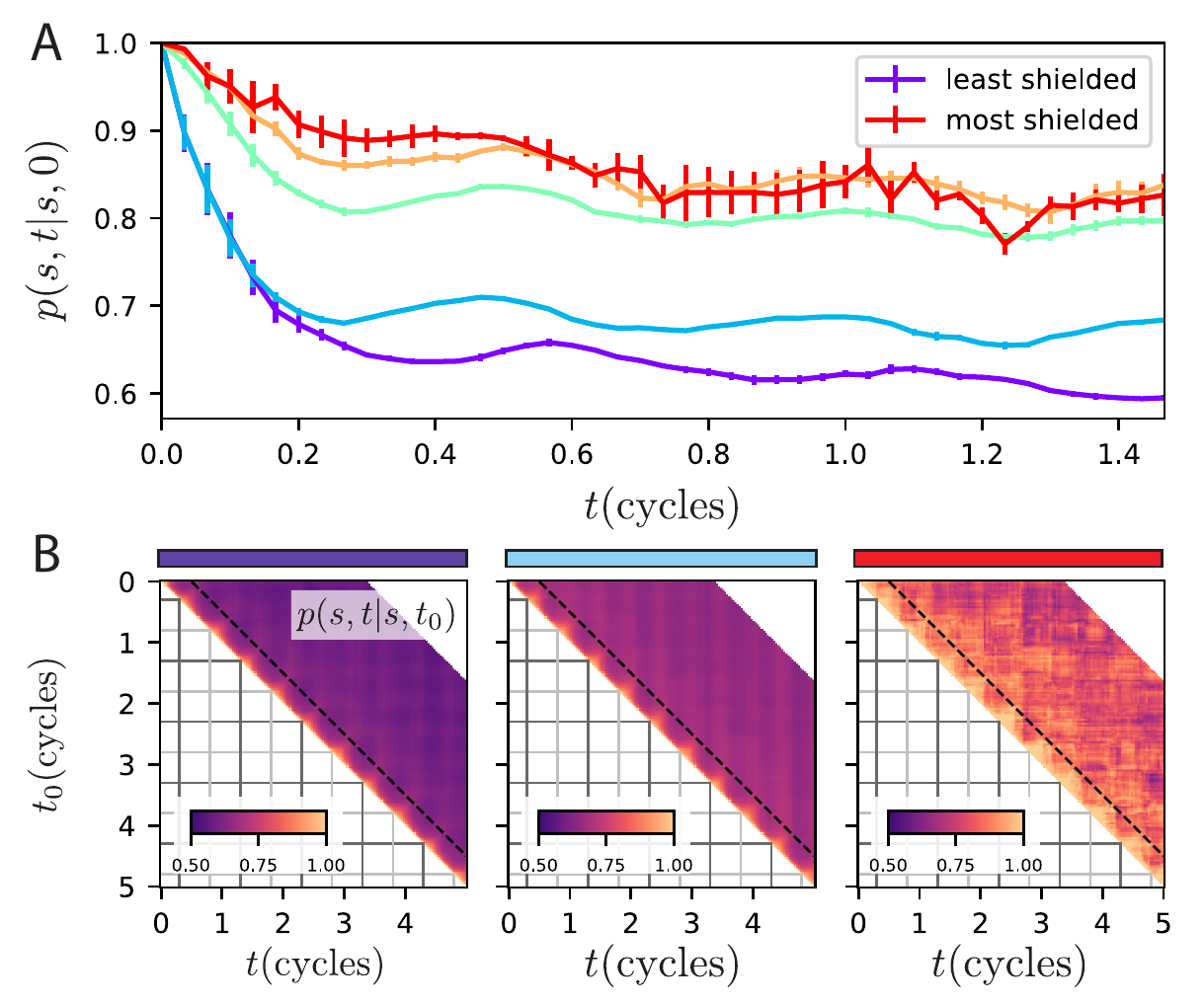}
\caption{\textbf{Structural correlations for the system at $\gamma_0 = 0.157$, the highest strain amplitude studied.}
(A) Structural auto-correlations $p(s,t \vert s, 0)$ as a function of $t$ over 1.5 shear cycles, with $t_0 = 0$ marking the beginning of the cycle.
Results are shown as stroboscopic averages, and error bars represent the standard error of the mean value at each time point.
Each signal represents a shielding level according to the color scheme detailed in Fig. \ref{fig:methods}, and is calculated over particles in the appropriate shielding level at $t_0$.
Colors of the least shielded (disordered) and most shielded layers are shown for reference.
(B) Full two-dimensional distributions of $p(s,t \vert s, t_0)$ for three shielding layers, shown beneath colored bars that indicate shielding level at time $t_0$ according to the color scheme detailed in Fig. \ref{fig:methods}.
Dark and light gray lines indicate times of strain extrema during each first and second half-cycle, respectively. 
Dotted lines indicate a one-dimensional slice through each distribution at $p(s, t \vert s, t - 0.5)$.
}
\label{fig:irrev}
\end{figure*}

We first address the system at the highest strain amplitude studied, $\gamma_0 = 0.157$, which displays per-particle structural correlations that are qualitatively different than structural correlations of systems at lower strain amplitudes (even when those systems are still above yield).
Fig. \ref{fig:irrev} displays some of these signals; together, they show that particles at $\gamma_0 = 0.157$ do not retain their distinct structural identities of crystalline or disordered in any significant capacity even over the course of one shear cycle.
The system is in a state of structural irreversibility on the individual particle level, even while structural reversibility exists at lower strain amplitudes that are still above yield.

Fig. \ref{fig:irrev}A shows $p(s,t \vert s, 0)$ for each shielding layer at $\gamma_0 = 0.157$ as a function of $t$ over 1.5 shear cycles, with $t_0 = 0$ marking the beginning of the cycle. 
This structural auto-correlation is calculated identically to that shown in Fig. \ref{fig:d2min}A for $\gamma_0 = 0.068$, still above the yield strain. 
Notably, whereas the correlations in Fig. \ref{fig:d2min}A appear periodic within one shear cycle and only show slow decay over longer timescales (Supplementary Figs. S10 and S12),
correlations in Fig. \ref{fig:irrev}A are never periodic and decay rapidly during the first cycle for all shielding layers, showing the destruction of maintained particle identities over the course of one cycle.
  
The decay in structural auto-correlation at $\gamma_0 = 0.157$ can be seen more fully in two-dimensional distributions of $p(s,t \vert s, t_0)$ across $t_0$ and $t$. Fig. \ref{fig:irrev}B shows distributions for particles in three crystalline shielding layers at time $t_0$: disordered particles, grain boundary particles, and particles that are most interior within crystal grains.
None of these two-dimensional distributions display meaningful periodicity over any one-dimensional cut through them, except the cut $p(s,t \vert s, t-0.5)$, drawn as a black dotted line.
This signal, explored thoroughly in the \emph{Results}, measures correlations only within the half-cycle time window, and indicates that the maximum period over which particles retain their structural identity is approximately half a cycle.

According to the oscillatory rheology shown in Fig. \ref{fig:methods}B, at the highest strain amplitude $\gamma_0 = 0.157$, the storage modulus $G'$ and loss modulus $G''$ are closest to each other, approaching equality.
We therefore posit that our particle-scale measurements of correlations in local structural homogeneity capture the macroscopic, rheological cross-over point at which material behavior is equally dominated by elastic and viscous response.

\subsection{Crystalline shielding in a mono-disperse system with larger crystal grains}
We also analyzed crystallinity and structural rearrangement in a mono-disperse system \cite{Keim2015}, identical in preparation to the bi-disperse one discussed in the \emph{Results}.
Due to the mono-dispersity, crystalline grains are larger, and thus crystalline shielding is deeper. Our results, shown in Supplementary Figs. S13, S14, and S15, are in agreement with those already presented. 
We find oscillating measures of global structural heterogeneity, both crystallinity $X(t)$ and $\overline{\Delta} (t)$ averaged over all particles, and an asymmetry in those measures with respect to shear direction as strain amplitude decreases, indicating that some material anisotropy and memory is preserved in the system (Supplementary Figs. S13A, S13B, S14).
Deep crystalline shielding exists in the system (Supplementary Figs. S13C, S13D, S13E), and particles of deeper shielding are generally less structurally volatile, with higher values of $p(s,t \vert s, t_0)$ (Supplementary Figs. S13F and S15).
Particles of deeper shielding are also generally more asymmetrical in their structural response with respect to shear direction: crystalline particles that are more shielded have probability signals $p(s,t \vert s, t-0.5)$ that are more dissimilar between each half of the shear cycle.
As a result, more shielded layers generally have lower values of $P_p (2\omega^*)/P_p (\omega^*)$ (Supplementary Fig. S13G).
These results are not as clear-cut as those obtained from the bi-disperse system, perhaps because in the mono-disperse system there are more crystalline layers, each with fewer particles, and statistics are consequently thinner. 
However, we can observe that the three most shielded layers have lower $P_p (2\omega^*)/P_p (\omega^*)$ values than the three least shielded layers at all strain amplitudes except for the highest strain amplitude studied. 
Particles of deeper shielding thus in general form regions in which the system's memory of its preparation is localized.
At the highest strain amplitude, we observe qualitatively distinct structural correlation measures across all shielding layers (Supplementary Fig. S15).
Correlations are not periodic even within one shear cycle, and differ in behavior from correlations at all lower strain amplitudes, in a similar manner to that discussed in Section \ref{section:irrev}.
Thus, particle structural identities are not retained even over one shear cycle.
The highest strain amplitude is again a point at which $G'$ and $G''$ approach equality as shown in Fig. \ref{fig:methods}B, implying that our local correlation measures capture a macroscopic rheological transition. 

\section{Conclusions}
In this paper, we have presented new and accessible measures on the individual particle level that correlate with non-trivial phenomena on the macroscale.
Our measures capture local structural homogeneity and its correlations over time, and are quite distinct from other quantities such as $D^2_{min}$ \cite{Falk1998} or T1 events \cite{Weaire2001} usually employed to investigate microstructural response under shear.
We have found that, by simply measuring correlations in the degree to which an individual particle's environment is similar to those of its neighbors, we can shed light on macroscale yield and memory effects in amorphous materials under oscillatory shear.

Our analysis indicates that the system-wide average of crystallinity on the particle level shows asymmetry with respect to shear direction, and thus encodes the material's memory of its preparation.
Correlations over time in the structural homogeneity of individual particles are reliable indicators of particle rearrangement as measured by $D^2_{min}$, and these correlations also show asymmetry with respect to shear direction.
The observed structural asymmetric response is not homogeneous throughout the system, however.
Particles that are more interior within crystal grains, or more ``shielded," are generally less structurally volatile over time, and intriguingly least symmetric in their structural response with respect to shear direction.
Thus, response asymmetry and consequent material memory is spatially localized in crystal grains.

Additionally, we have found that structural correlations are qualitatively different at strain amplitudes for which macroscale rheological measures of elastic and viscous response approach cross-over, implying that our local measurements indicate a behavioral transition usually only visible on a much larger length scale.
Our work bridges the micro- and macro- scales, and thus will be useful for future experimentalists studying yield in amorphous systems who may have access only to information on one length scale, either microstructural or rheological.
Our efforts add to the growing body of knowledge regarding the nature of microscopic rearrangement and macroscopic yield in disordered materials, and help to illuminate how, and specifically where, certain types of memory are stored in these systems.

\section{Citation diversity statement}
Recent work in several fields of science has identified a bias in citation practices such that papers from women and other minorities are under-cited relative to other papers in the field \cite{Dworkin2020, maliniak2013gender, caplar2017quantitative, chakravartty2018communicationsowhite, YannikThiemKrisF.SealeyAmyE.FerrerAdrielM.Trott2018, dion2018gendered}. 
Here we sought to proactively consider choosing references that reflect the diversity of our field in thought, form of contribution, gender, and other factors.
We obtained predicted gender of the first and last author of each reference by using databases that store the probability of a name being carried by a woman or a man \cite{Dworkin2020,cleanbib}; we supplemented these results with online research of individuals for whom automatic classification failed. By this measure (and excluding self-citations to the first and last authors of our current paper), our references contain $70.5\%$ man/man, $15.9\%$ man/woman, $9.1\%$ woman/man, and $4.5\%$ woman/woman categorization. The automated method is limited in that a) names, pronouns, and social media profiles used to construct the databases may not, in every case, be indicative of gender identity and b) it cannot account for intersex, non-binary, or transgender people. We look forward to future work that could help us to better understand how to support equitable practices in science. 

\section{Acknowledgements}
We thank Michael Engel for use of the software package inJaVis, used to visualize the particle systems.
E.G.T, K.L.G, P.E.A, and D.S.B. are supported by the National Science Foundation Materials Research Science and Engineering Center at University of Pennsylvania (NSF grant DMR-1120901).
K.L.G. and P.E.A are additionally supported by the U.S. Army Research Office (ARO grant W911-NF-16-1-0290).
E.G.T and D.S.B. are additionally supported by the Paul G. Allen Family Foundation.
The views and conclusions contained in this document are solely those of the authors.

\bibliographystyle{apsrev4-2}
\bibliography{dsb_mat,misc,diversity}

\end{document}


\title{Supplementary Information: Crystalline shielding mitigates structural rearrangement and localizes memory in jammed systems under oscillatory shear}

\author{Erin G. Teich}
\affiliation{Department of Bioengineering, University of Pennsylvania, Philadelphia PA 19104}
\author{K. Lawrence Galloway}
\affiliation{Department of Mechanical Engineering and Applied Mechanics, University of Pennsylvania, Philadelphia PA 19104}
\author{Paulo E. Arratia}
\affiliation{Department of Mechanical Engineering and Applied Mechanics, University of Pennsylvania, Philadelphia PA 19104}
\affiliation{Department of Chemical and Biomolecular Engineering, University of Pennsylvania, Philadelphia PA 19104}
\author{Danielle S. Bassett}
\affiliation{Department of Bioengineering, University of Pennsylvania, Philadelphia PA 19104}
\affiliation{Department of Physics and Astronomy, University of Pennsylvania, Philadelphia PA 19104}
\affiliation{Department of Electrical and Systems Engineering, University of Pennsylvania, Philadelphia PA 19104}
\affiliation{Santa Fe Institute, Santa Fe NM 87501}

\date{\today}

\maketitle

\section{Continuous per-particle structural homogeneity}
\begin{figure*}
\centering
\includegraphics[width=\textwidth]{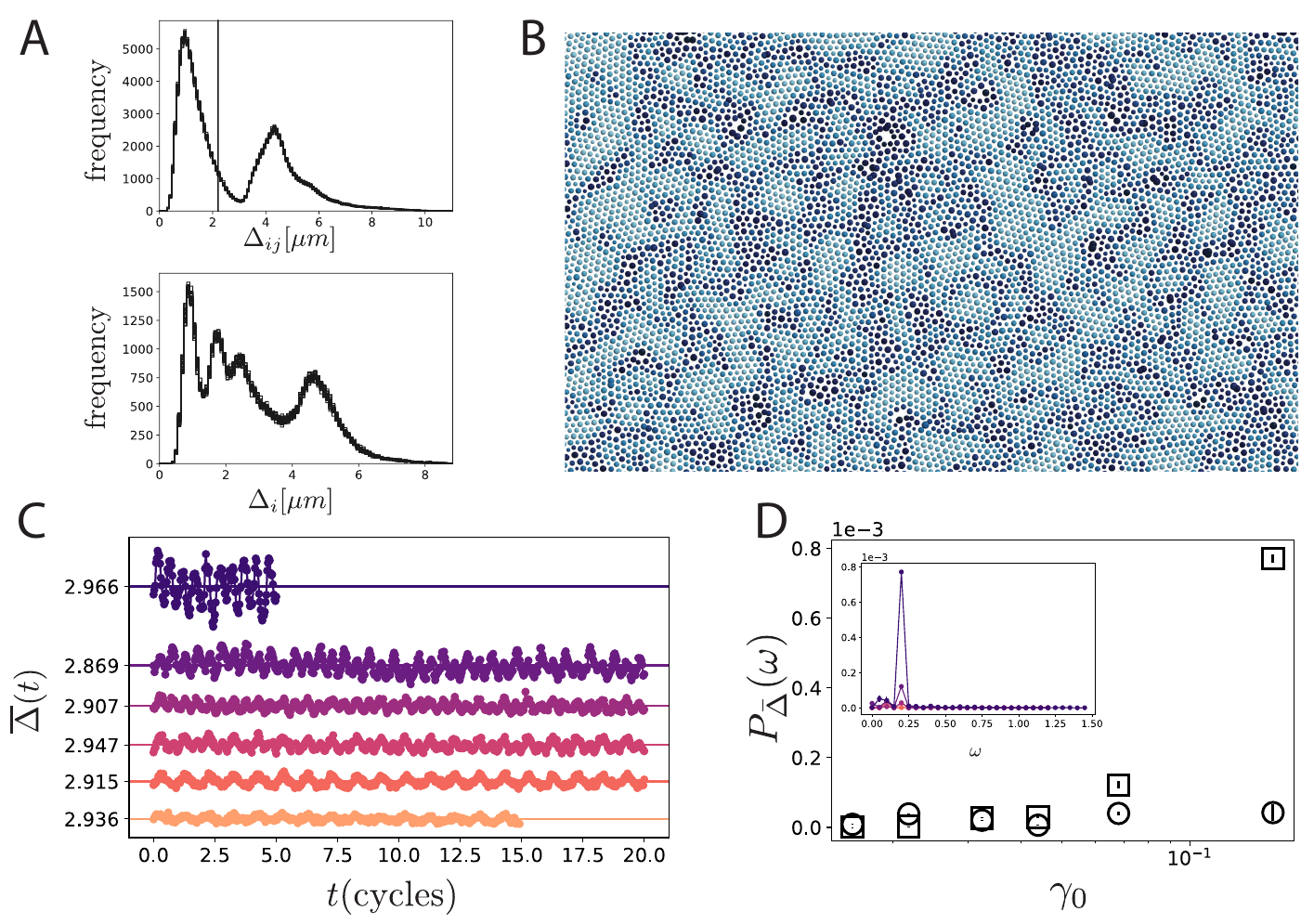}
\caption{
Continuous per-particle structural homogeneity also shows asymmetry with respect to shear direction.
(A) Overlaid histograms of $\Delta_{ij}$ over all neighbor pairs $ij$ (top panel) in 2 cycles of a sample experiment at $\gamma_0 = 0.068$, and $\Delta_i$ over all particles (bottom panel) in the same 2 cycles.
In the top panel, a vertical line marks the threshold $t$ used to determine crystallinity as described in the \emph{Methods} section of the main text.
(B) Visual rendering of a portion of the system at $\gamma_0 = 0.068$ with particles colored according to $\Delta_i$. Lighter blue particles have lower values of $\Delta_i$, and thus exhibit a higher crystalline quality. 
(C) The quantity $\overline{\Delta} (t)$ in $\mu m$, averaged over all particles in each system, as a function of strain amplitude $\gamma_0$. Signatures are arbitrarily offset for clarity; higher curves, with darker colors, correspond to higher values of $\gamma_0$. Horizontal lines indicate the mean of each signature.
(D) Two power spectral densities $P_{\bar{\Delta}} (\omega)$ of each $\overline{\Delta} (t)$ signature as a function of $\gamma_0$, for two distinct frequencies. Circles correspond to frequency $\omega^*$, the frequency of the needle oscillation, and squares correspond to frequency $2\omega^*$, or the second harmonic of the needle oscillation. 
Error bar estimation is detailed in the supplementary text.
Inset are full power spectral densities $P_{\bar{\Delta}} (\omega)$ for all $\omega$ at all strain amplitudes.
}
\label{fig:methods_RMSD}
\end{figure*}

We may also characterize structural homogeneity continuously, by measuring the root-mean-squared deviation, $\Delta_{ij}$, between the environments of particles $i$ and $j$ ($\{ \bm{r}_{im} \}$ and $\{ \bm{r}_{jm'} \}$ respectively, defined in the main text):
\begin{align}
\Delta_{ij} = \sqrt{\frac{1}{M} \sum_{m,m'}(\bm{r}_{im} - \bm{r}_{jm'})^2}.
\end{align}
The sum proceeds over the mapping $(m,m')$ found that best minimizes $\Delta_{ij}$.
To obtain this mapping, we consider each vector in the set $\{ \bm{r}_{jm'} \}$ in turn, and greedily pair it with the closest vector in the set $\{ \bm{r}_{im} \}$ that is unpaired.
(We note that the mapping found this way is not guaranteed to give the global $\Delta_{ij}$ minimum, since finding the global minimum amounts to solving the well-known assignment problem \cite{Aardal2005} and we did not implement any of its algorithmic solutions.)
When computing $\Delta_{ij}$, particles $i$ and $j$ are allowed to have different numbers of nearest neighbors; we set $M$ to be the larger of the pair $[M_i, M_j]$ and simply augment the smaller set of environment vectors with $\bm{0}$ vectors until both sets are the same size.
Example distributions of $\Delta_{ij}$ over all neighbor pairs $ij$ in a sample experiment are shown in Fig. \ref{fig:methods_RMSD}A, with the threshold $t$ used to determine crystallinity in the main text also shown for reference. 
We set $\Delta_{ij} = \Delta_{ji}$ for computational efficiency when computing this quantity over all neighbor pairs; this equality always holds if $\Delta_{ij}$ is a true global minimum.

The set of pairwise values $\Delta_{ij}$ for each particle $i$ can be averaged over all of its neighbors $j$ to produce a per-particle continuous measure of structural homogeneity, which we simply call $\Delta_i$. This measure is lower for less disordered (more structurally homogeneous, or more crystalline) particles, and higher for more disordered particles. Example distributions of $\Delta_i$ over all particles in a sample experiment are shown in Fig. \ref{fig:methods_RMSD}A, and a snapshot of an experimental system colored by this structural order parameter is shown in Fig. \ref{fig:methods_RMSD}B.
Global structural homogeneity can then be defined as $\overline{\Delta}(t)$, where the average is taken over $\Delta_i$ for all particles in the system at each time.
Figs. \ref{fig:methods_RMSD}C and \ref{fig:methods_RMSD}D show that global structural homogeneity defined in this manner exhibits similar asymmetry to the global crystallinity explored in the main text.

\subsection{Differences in structural homogeneity over time indicate structural rearrangement under shear}
\begin{figure*}
\centering
\includegraphics[width=\textwidth]{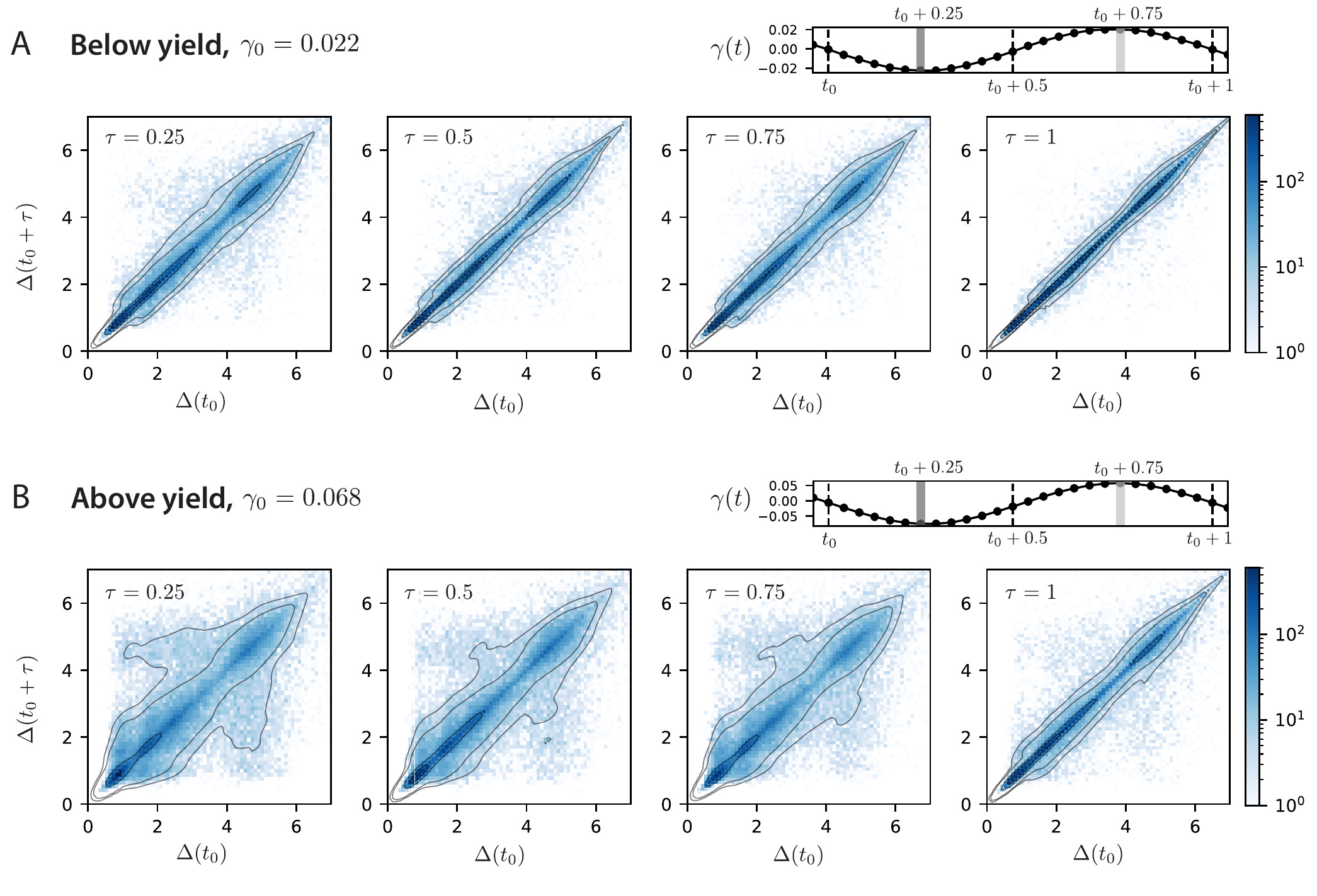}
\caption{
Joint histograms of $\Delta$ in $\mu m$ at the start $t_0$ of a shear cycle and at various points $t_0 + \tau$ during the cycle for example systems (A) below and (B) above yield. 
Strain $\gamma (t)$ for the cycle is shown above each set of histograms, with lines and labels indicating times of interest. 
Joint histograms correspond to $t_0$ against each of these times according to the label in their upper left corners, and proceed chronologically from left to right.
Histograms are colored logarithmically according to color bars displayed to the right of each set of panels.
}
\label{fig:RMSD}
\end{figure*}

A closer investigation of the structural homogeneity of individual particles over time reveals that differences in homogeneity at various points during the shear cycle are reliable indicators of structural rearrangement.
We track how ensemble distributions of $\Delta_i$ evolve over the course of one shear cycle at two strain amplitudes: one below the yield strain (Fig. \ref{fig:RMSD}A) and one above it (Fig. \ref{fig:RMSD}B).
We find that within-cycle structural rearrangement exists both above and below the yield strain amplitude, and between-cycle rearrangement exists significantly only above yield.
Our results agree with previous work \cite{Keim2014, Keim2015} that found reversible (within-cycle) particle rearrangements both above and below yield, and increasing numbers of irreversible (between-cycle) particle rearrangements in these systems as strain amplitude increases past the yield strain. 

Fig. \ref{fig:RMSD} shows joint distributions of $\Delta(t_0)$ at the start of a typical shear cycle (well after any transient behavior has died off) and $\Delta(t_0+\tau)$ at various subsequent points of the shear cycle.
To obtain contour lines, joint distributions were first smoothed via kernel density estimation with a Gaussian kernel whose bandwidth was set to be 1.5 times the bin width of the joint histogram. 
Contour lines were then drawn at levels [5, 10, 100, 200] of the smoothed histogram.
Diagonal elements of these distributions represent particles that have the same structural homogeneity at $t_0$ and $t_0+\tau$, while off-diagonal elements represent particles for which there is significant discrepancy between structural homogeneity at $t_0$ and that at $t_0+\tau$. Off-diagonal elements therefore represent particles that have experienced structural rearrangement from time $t_0$ to $t_0+\tau$. 
We first note that structural rearrangement is more evident for the system above yield than for the system below yield, since the joint $\Delta$ distribution above yield contains generally more off-diagonal elements than the corresponding distribution below yield at all lag times $\tau$.
Both above and below yield, within-cycle structural rearrangement occurs, evidenced by off-diagonal elements in the joint $\Delta$ distribution at $t_0$ and $t_0+\tau$ when $\tau < 1$ cycle.
Rearrangement can be seen especially for $t_0$ at the beginning of the shear cycle and $t_0+\tau$ at subsequent strain extrema ($\tau = 0.25, 0.75$ cycles). 
Evidently throughout these windows particles undergo structural rearrangement either from a more homogeneous local environment at time $t_0$ to a disordered one at time $t_0+\tau$, or \emph{vice versa}.
The system above yield also shows between-cycle structural rearrangement, as the joint $\Delta$ distribution at $t_0$ and $t_0+1$ cycles contains off-diagonal elements. 
The system below yield shows fewer between-cycle structural rearrangements and its corresponding joint distribution hews more closely to the diagonal.

Both above and below yield, structural rearrangement is more evident during the major shear half-cycle than during the minor shear half-cycle: joint distributions of $\Delta$ for $t_0$ and $t_0+0.25$ cycles are more distributed off the diagonal than those for $t_0$ and $t_0+0.75$ cycles.
This asymmetry is additional evidence of a difference in structural response according to shear direction, as explored thoroughly in the main text.

\begin{figure*}
\centering
\includegraphics[width=0.6\textwidth]{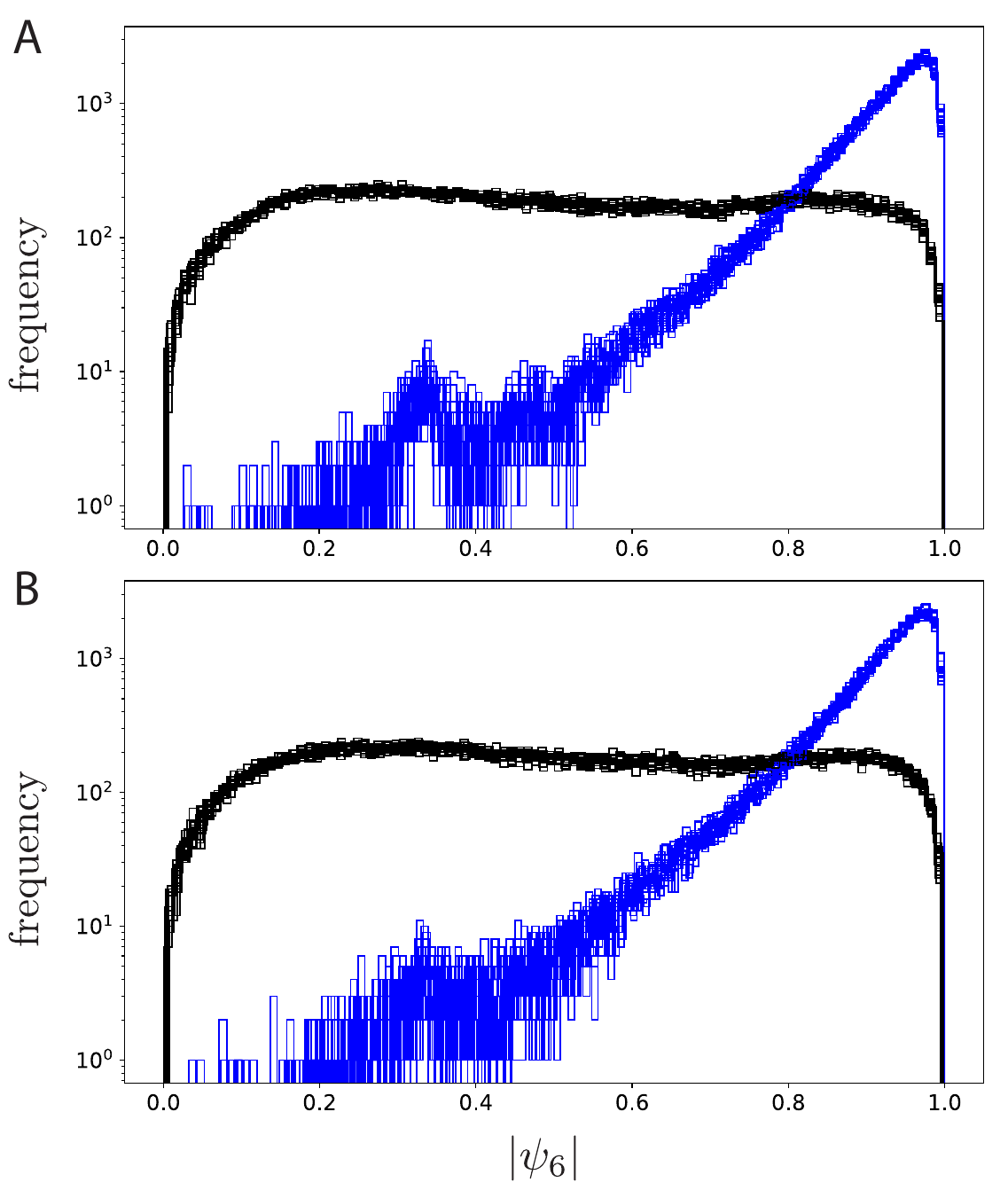}
\caption{
Distributions of the bond orientational order parameter $\vert \psi_6 \vert$, collected separately for crystalline particles (blue) and non-crystalline particles (black) over one cycle, for experiments (A) below and (B) above yield.
Experiments are at $\gamma_0 = 0.022$ and $\gamma_0 = 0.068$, respectively.
The quantity $\vert \psi_6 \vert$ for each particle $i$ is the absolute value of the complex number $\psi_6 = \frac{1}{N_i} \sum_{j=1}^{N_i} e^{6i\phi_{ij}}$ as defined in the main text.
In these calculations, we set $N_i = 6$ for all particles.
}
\label{fig:hexatic}
\end{figure*}

\begin{figure*}
\centering
\includegraphics[width=\textwidth]{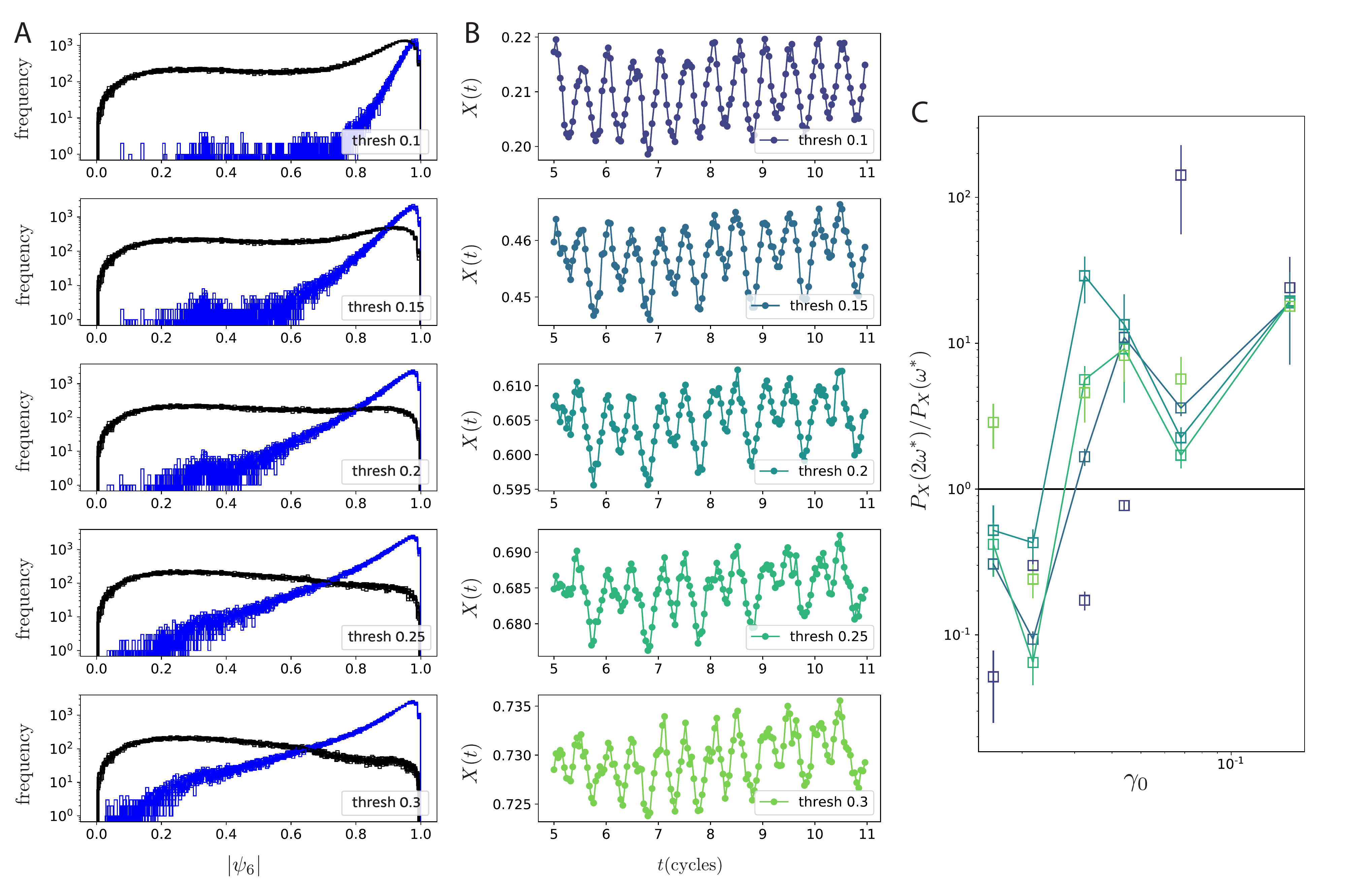}
\caption{
The influence of matching threshold on crystallinity characterization.
(A) Distributions of the bond orientational order parameter $\vert \psi_6 \vert$, collected separately for crystalline particles (blue) and non-crystalline particles (black) over one cycle, for an experiment above yield at $\gamma_0 = 0.068$. 
Each panel corresponds to a different matching threshold, labeled with thresh  = $t / r_{cut}$ as defined in the main text.
At the most stringent $t=0.1r_{cut}$, a significant number of non-crystalline particles have high values of $\vert \psi_6 \vert$, while at the most lenient $t=0.3r_{cut}$, a significant number of crystalline particles have low values of $\vert \psi_6 \vert$. 
The threshold $t=0.2r_{cut}$ thus produces a bipartition that is a reasonable compromise between these two extremes.
(B) Global crystallinity for the same experiment, calculated according to various thresholds over a number of cycles in the non-transient regime.
Although the mean value of crystallinity varies widely across threshold, from $\sim 0.21$ at $t=0.1r_{cut}$ to $\sim 0.73$ at $t=0.3r_{cut}$, crystallinity shows qualitatively similar oscillatory behavior in all cases.
(C) The power spectral density of global crystallinity at $2 \omega^*$ (twice the frequency of the shear cycle), normalized by the power spectral density of global crystallinity at $\omega^*$, for varying thresholds. 
Signals are shown as a function of strain amplitude $\gamma_0$ and colored according to threshold as shown in panel B.
Note that, although the choice of threshold significantly influences the identification of crystalline particles as shown in panel A, the choice of threshold (within reason) does not influence one of the main results of the paper, that crystallinity shows asymmetry with respect to shear direction at strain amplitudes below yield.
For the first two strain amplitudes (below yield), the ratio of power spectral densities is below 1 for $t=0.15r_{cut}$, $t=0.2r_{cut}$, and $t=0.25r_{cut}$, indicating that the crystallinity is dominated by the first harmonic of the shear cycle, and is thus asymmetric with respect to shear direction.
For subsequent strain amplitudes (above yield), the ratio of power spectral densities is above 1 for these thresholds, indicating that the crystallinity is more dominated by the second harmonic of the shear cycle, and is thus more symmetric with respect to shear direction.
Each power spectral density, calculated via the periodogram estimate, is the mean of a set of $P_X(\omega)$ values calculated over consecutive 2 cycle windows of 6 cycles in the non-transient regime of each experiment.
Error bars of the displayed power spectral density ratio are calculated via Taylor series propagation of the standard errors of the mean of each $P_X(\omega)$ quantity in the ratio.
}
\label{fig:threshold}
\end{figure*}

\begin{figure*}
\centering
\includegraphics[width=0.6\textwidth]{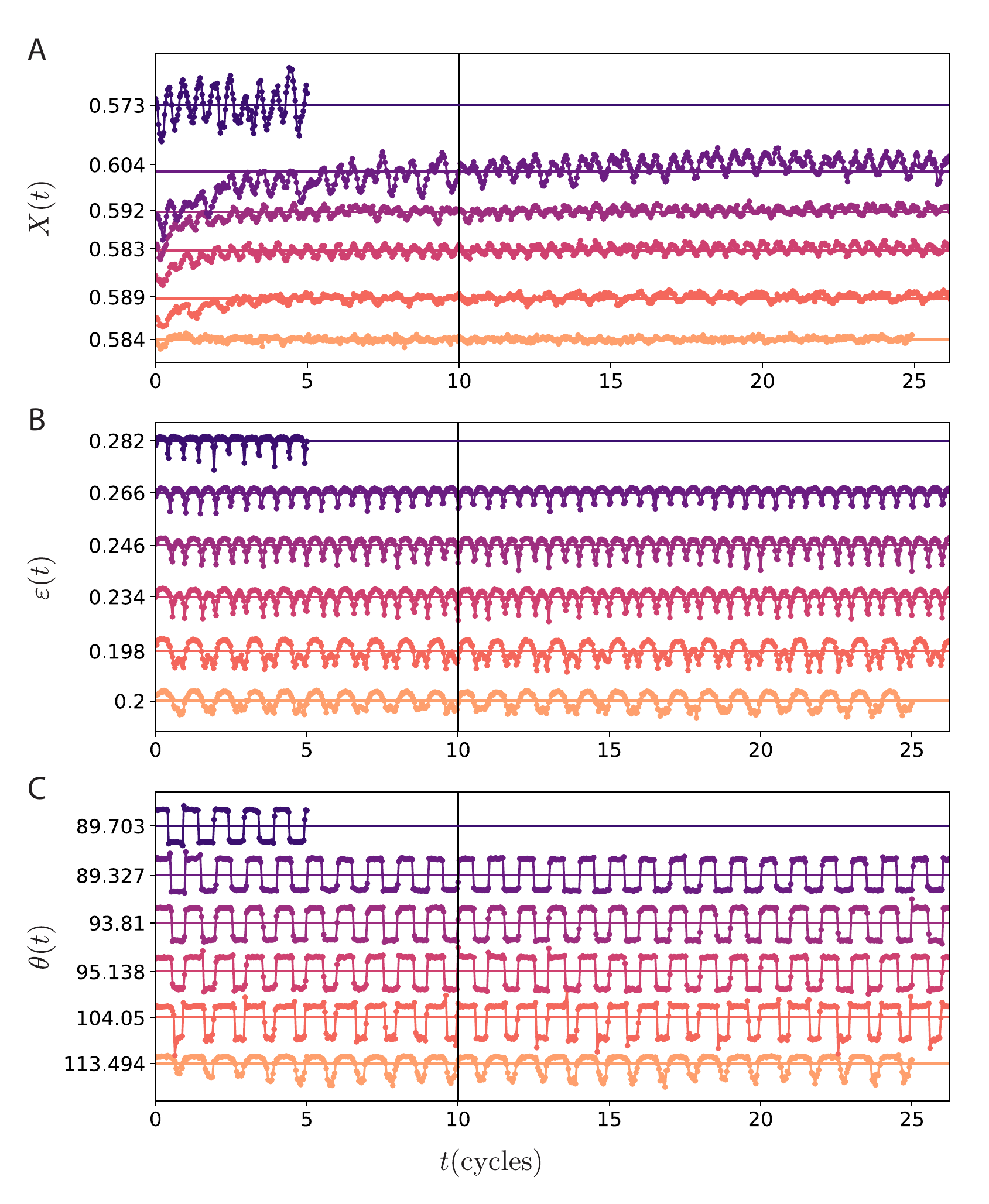}
\caption{
Structural signals (A) crystallinity, (B) local neighborhood ellipse eccentricity, and (C) local neighborhood ellipse orientation, for entire trajectories at all strain amplitudes. 
Signatures are arbitrarily offset for clarity; higher curves, with darker colors, correspond to higher values of $\gamma_0$.
Each panel shows a vertical line after which we define the system trajectory to be non-transient.
The exception is the system at the highest strain amplitude studied, for which we define the entire trajectory shown to be non-transient.
Horizontal lines in each panel indicate the mean of each non-transient signature.
}
\label{fig:transient}
\end{figure*}

\begin{figure*}
\centering
\includegraphics[width=0.8\textwidth]{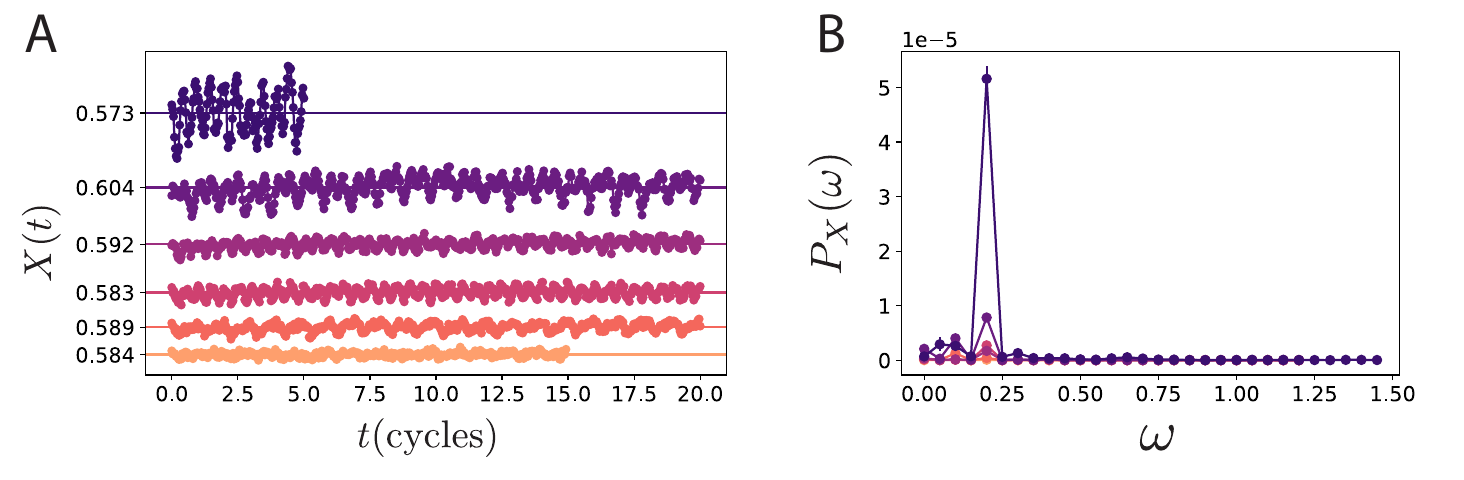}
\caption{
Global crystallinity for whole non-transient trajectories as a function of strain amplitude $\gamma_0$.
(A) Crystallinity in all systems. Signatures are arbitrarily offset for clarity; higher curves, with darker colors, correspond to higher values of $\gamma_0$. Horizontal lines indicate the mean of each crystallinity signature.
(B) Power spectral densities $P_X(\omega)$ of each crystallinity signature at all strain amplitudes.
Error bar estimation is detailed in the main text.
}
\label{fig:xtal_full}
\end{figure*}

\begin{figure*}
\centering
\includegraphics[width=0.5\textwidth]{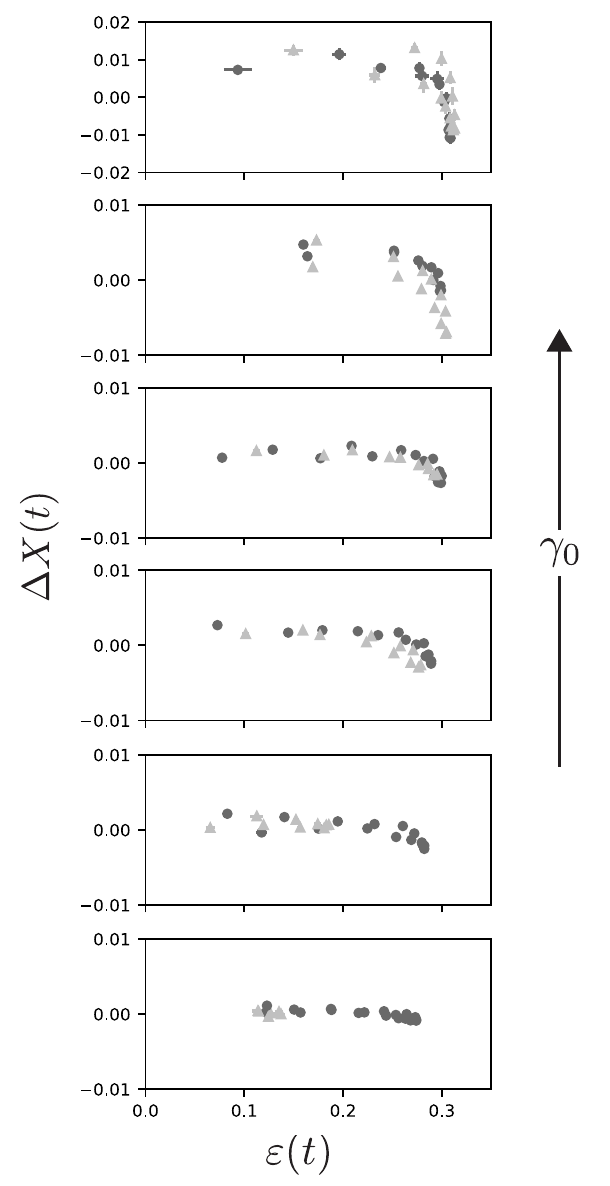}
\caption{
Deviations in the crystallinity from its mean, $\Delta X(t) \equiv X(t) - \langle X \rangle_t$, plotted against local neighborhood deformation $\varepsilon (t)$ for all systems. 
Shown are stroboscopically averaged quantities and corresponding error bars as detailed in the main text.
Systems are ordered in increasing strain amplitude $\gamma_0$ as indicated by the arrow.
Light gray triangles mark all frames for which $\theta(t) \leq 90^\circ$, and dark gray circles mark all frames for which $\theta(t) > 90^\circ$.
}
\label{fig:xtal_ecc}
\end{figure*}

\begin{figure*}
\centering
\includegraphics[width=\textwidth]{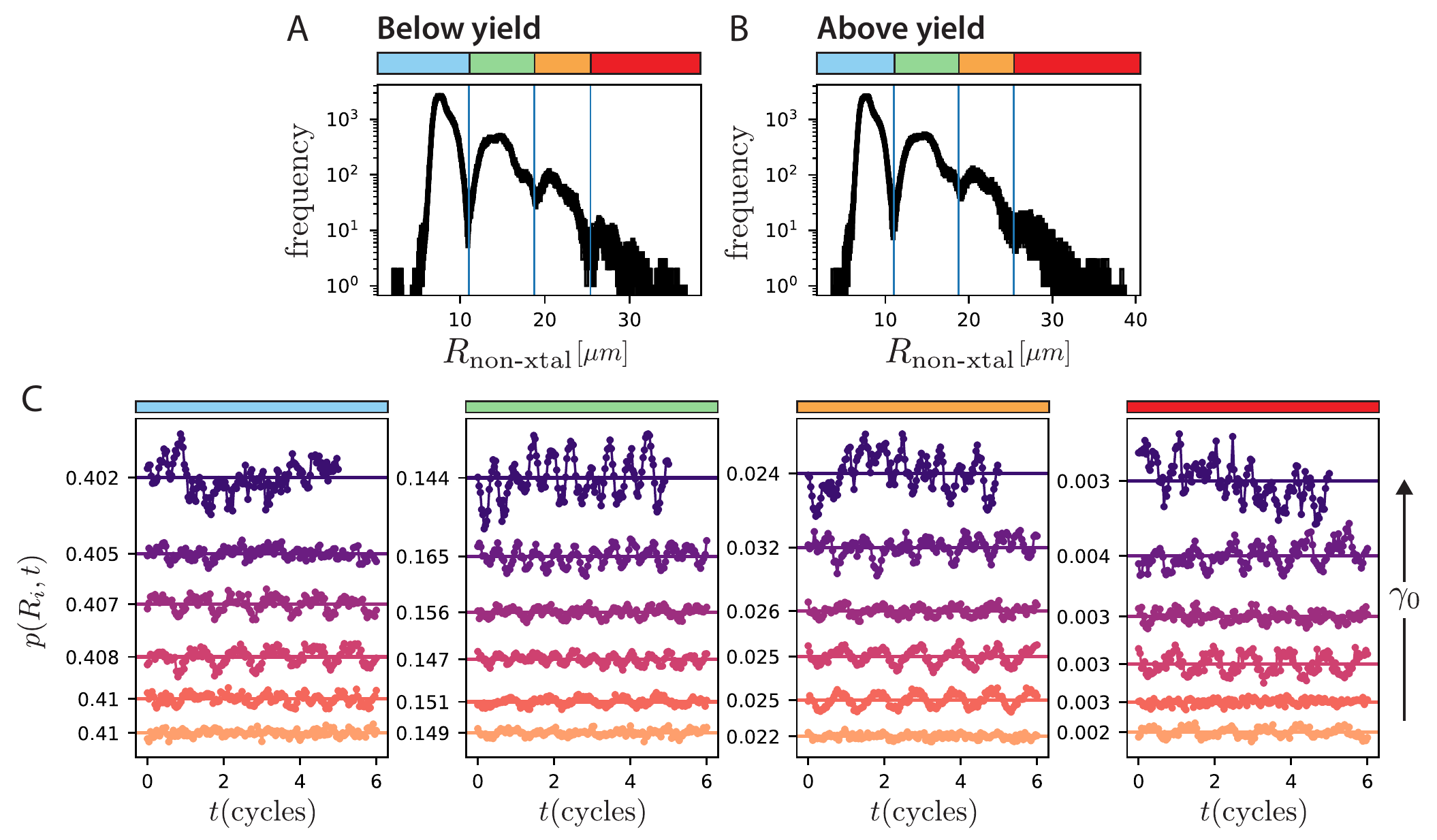}
\caption{Crystalline particles at different levels of ``shielding," or interiority within crystal grains, have varying responses to oscillatory shear.  
(A,B) Overlaid histograms of $R_{\text{non-xtal}}$ in $\mu m$ for all particles in 6 consecutive non-transient cycles for experiments (A) below and (B) above yield.
Experiments are at $\gamma_0 = 0.022$ and $\gamma_0 = 0.068$, respectively.
(C) The fraction of particles at each shielding level, $p(R_i, t)$, over a sample number of cycles. Signatures of experiments at higher $\gamma_0$ are colored darker. 
Particle fractions at all shielding levels generally exhibit oscillations of larger amplitude as strain amplitude increases.
Mean values for each signature are expressed with horizontal lines.
Signatures are shown beneath colored bars that indicate shielding level according to the color scheme detailed in panels A,B.
}
\label{fig:protection}
\end{figure*}

\begin{figure*}
\centering
\includegraphics[width=0.4\textwidth]{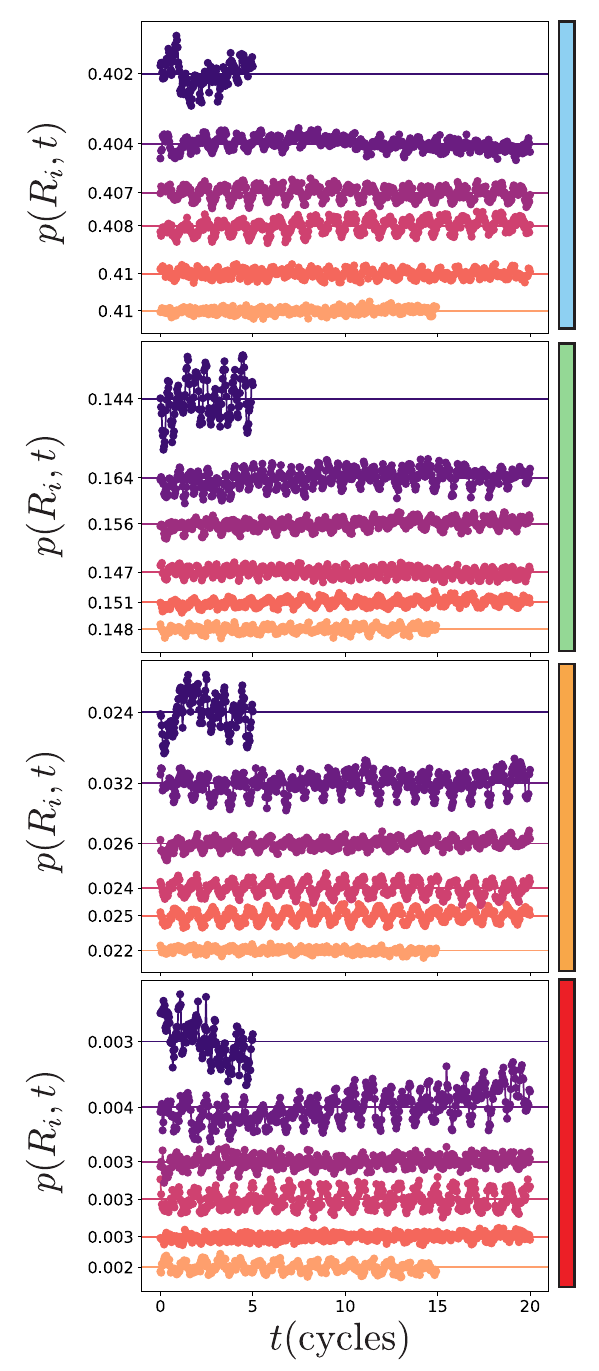}
\caption{The fraction of particles at each shielding level, $p(R_i, t)$, for whole non-transient trajectories. Signatures of experiments at higher $\gamma_0$ are colored darker. Mean values for each signature are expressed with horizontal lines.}
\label{fig:xtal_Rslice_full}
\end{figure*}

\begin{figure*}
\centering
\includegraphics[width=0.9\textwidth]{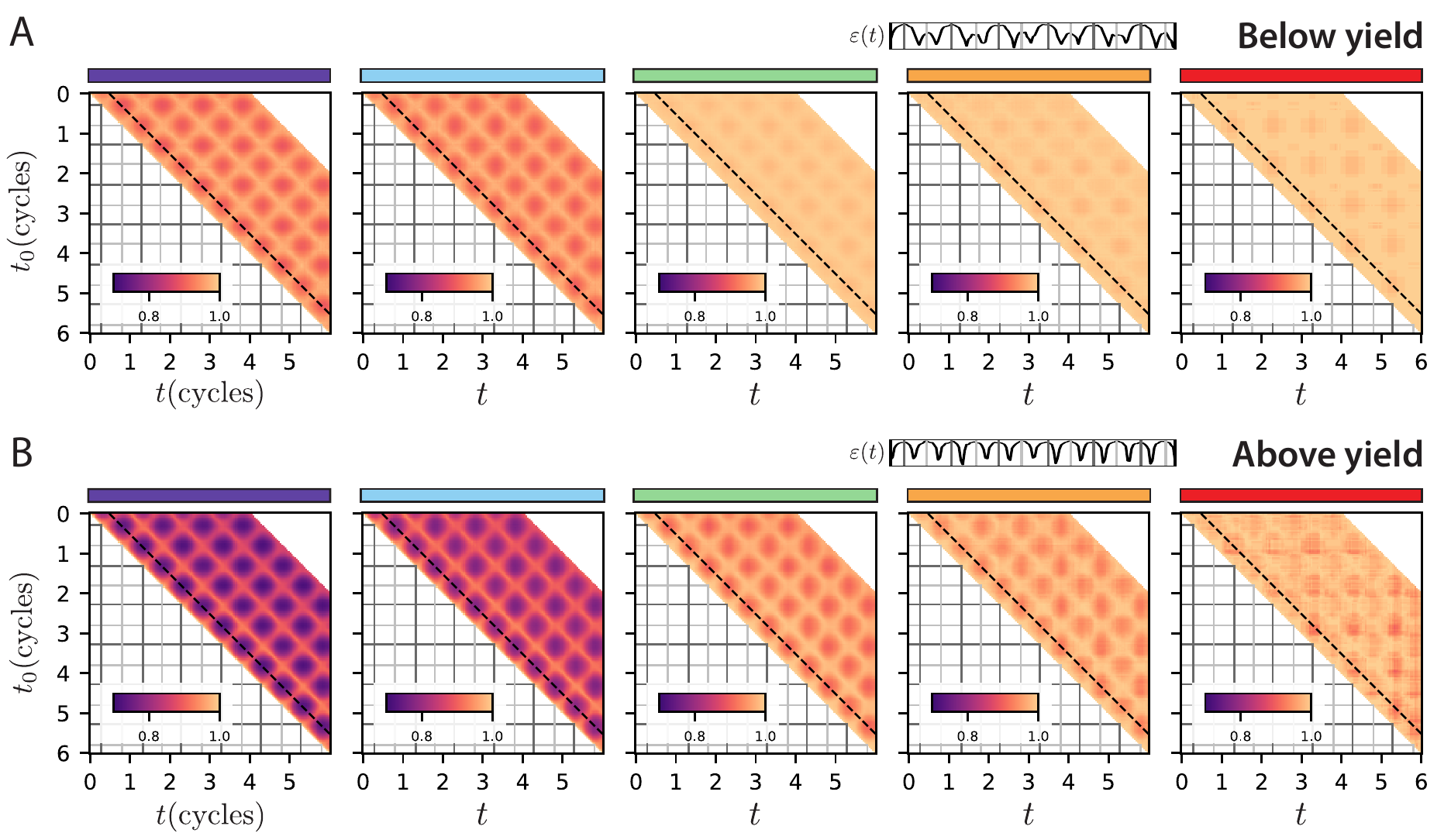}
\caption{Structural auto-correlations $p(s,t \vert s,t_0)$ for disordered and crystalline particles of different shielding levels in experiments (A) below and (B) above yield.
Experiments are at $\gamma_0 = 0.022$ and $\gamma_0 = 0.068$, respectively.
Auto-correlations are shown beneath colored bars that indicate shielding level of the particles at time $t_0$.
Distributions contain results for all time windows satisfying $(t_0 \leq t \leq t_0 + 4$ cycles) for a sample cycle set; results for full non-transient trajectories are shown in Fig. \ref{fig:xcorr_full}.
Dotted lines indicate a one-dimensional slice through each distribution at $p(s, t \vert s, t - 0.5)$.
Dark and light gray lines indicate times of first half-cycle and second half-cycle strain extrema, respectively, and local deformation at these extrema is shown via the plot of neighborhood ellipse eccentricity over time above each panel. 
Auto-correlations are oscillatory at all shielding levels, displaying clear minima when $t_0$ and $t$ are opposite strain extrema.
Note that auto-correlations for particles of lower crystalline shielding generally reach deeper minima by eye, and auto-correlations for particles of any shielding level in the system above yield reach deeper minima than those for the corresponding system below yield.
} 
\label{fig:corr}
\end{figure*}

\begin{figure*}
\centering
\includegraphics[width=0.9\textwidth]{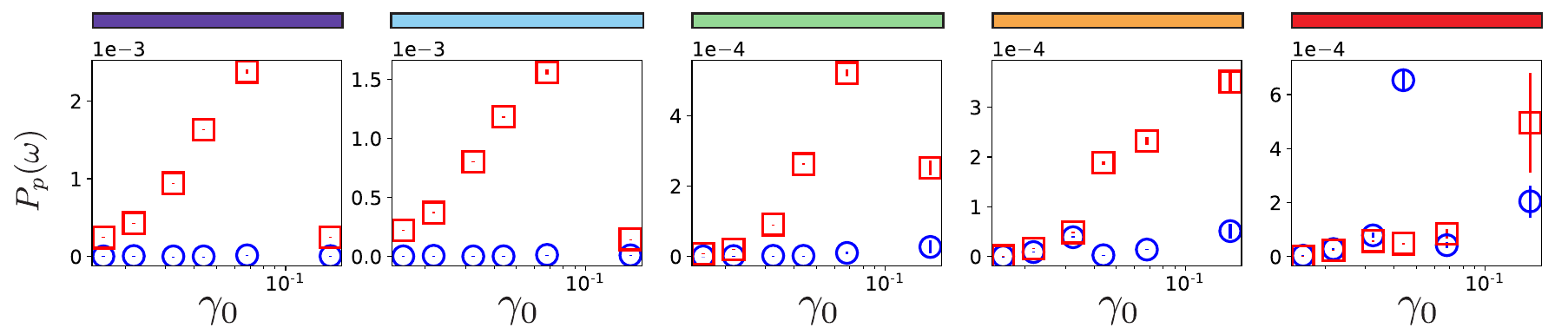}
\caption{Two power spectral densities, $P_p(\omega)$, of each crystallinity signature $p(s, t \vert s, t-0.5)$ as a function of $\gamma_0$ at each shielding level. 
Blue circles correspond to frequency $\omega^*$, the frequency of the needle oscillation, and red squares correspond to frequency $2\omega^*$, or the second harmonic of the needle oscillation. 
Each value is the mean of a set of $P_p(\omega)$ values calculated over consecutive 2 cycle windows of the full $p(s, t \vert s, t-0.5)$ signal shown in Fig. \ref{fig:xcorr_full}, 
and error bars are standard deviations of the mean.
There is a growth in $P_p (2\omega^*)$ as strain amplitude increases for all but the most shielded crystalline layer, indicating that particles in most shielding layers are increasingly responsive during the second shear half-cycle. 
For the most shielded layer, $P_p (2\omega^*)$ does not grow to dominate $P_p (\omega^*)$.
Note that patterns of systems at lower strain amplitudes are not maintained for the highest strain amplitude studied, due to the significant and fast decay of all correlation signatures for all shielding layers.
} 
\label{fig:FFTcorr}
\end{figure*}

\begin{figure*}
\centering
\includegraphics[width=0.9\textwidth]{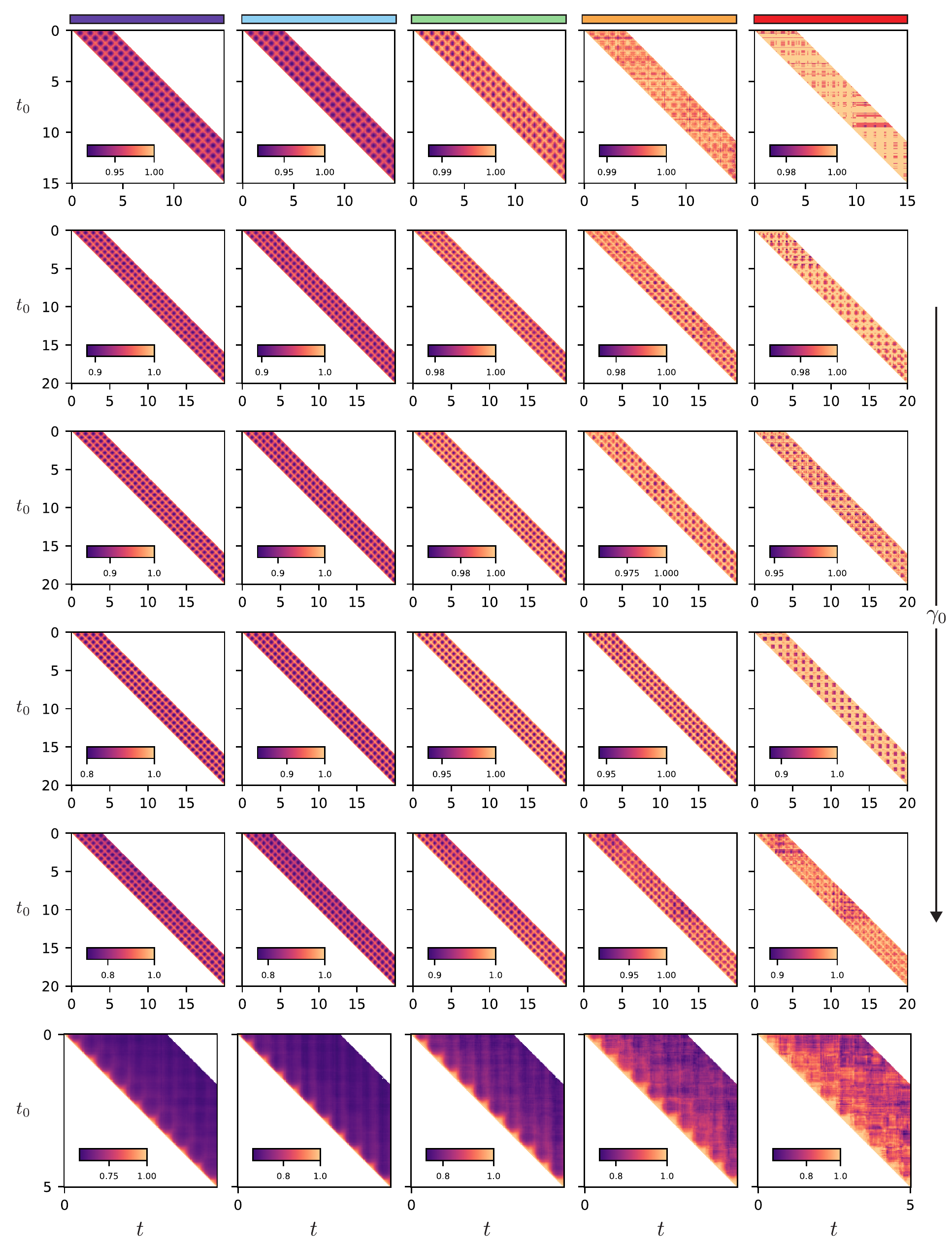}
\caption{Distributions of $p(s,t \vert s, t_0)$, with colored bars above each plot indicating the shielding level, for all systems.
Systems are ordered in increasing strain amplitude $\gamma_0$ as indicated by the arrow.
Signals are only shown for the non-transient portions of the experimental trajectories.
}
\label{fig:xcorr_full}
\end{figure*}

\begin{figure*}
\centering
\includegraphics[width=0.9\textwidth]{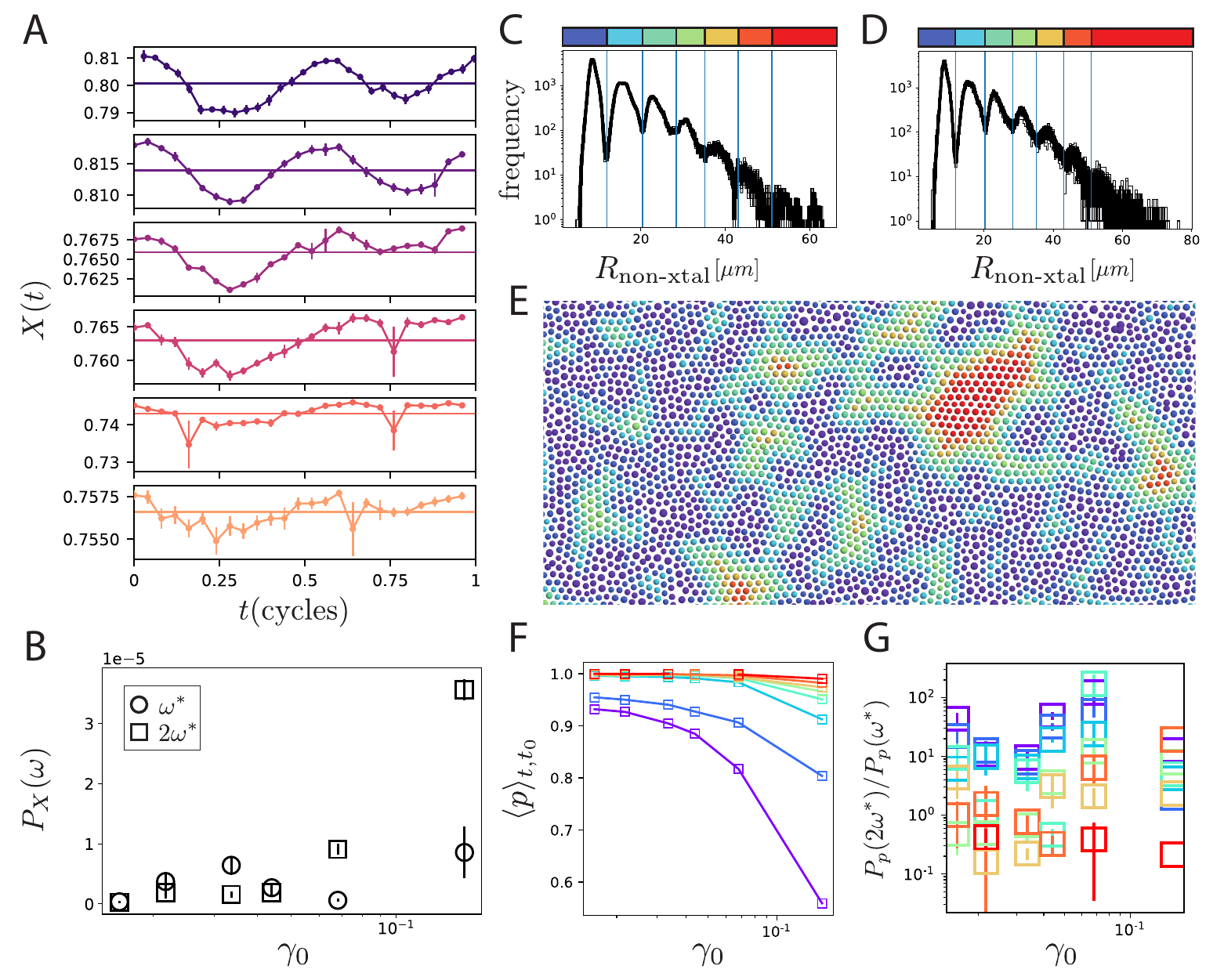}
\caption{Summary of results for mono-disperse data.
(A) Crystallinity in all systems, stroboscopically averaged. 
Signatures with darker colors correspond to higher values of $\gamma_0$. 
Horizontal lines indicate the mean of each signature. 
(B) Two power spectral densities $P_X(\omega)$ of each crystallinity signature as a function of $\gamma_0$, for two distinct frequencies. Circles correspond to frequency $\omega^*$, the frequency of the needle oscillation, and squares correspond to frequency $2\omega^*$, or the second harmonic of the needle oscillation. 
Each power spectral density is the mean of a set of $P_X(\omega)$ values calculated over consecutive $n$ cycle windows of the full non-transient relevant trajectory shown in Fig. \ref{fig:mono_xtal_full}, and error bars represent the standard deviation of the mean. We set $n = (5,7,7,7,7,2)$ for systems ordered by increasing strain amplitude.
(C,D) Overlaid histograms of $R_{\text{non-xtal}}$ in $\mu m$ for experiments (C) below and (D) above yield over 6 sample cycles.
Experiments are at $\gamma_0 = 0.019$ and $\gamma_0 = 0.060$, respectively.
(E) Rendering of an example system, with particle radii drawn as twice the measured image radii of gyration reported in Ref. \citenum{Keim2015} and particles colored according to crystalline shielding.
Disordered particles are colored purple.
(F) Average $\langle p(s,t \vert s, t_0) \rangle_{t_0,t}$ at all shielding levels as a function of strain amplitude $\gamma_0$.
Signals are colored by shielding level according to the color scheme detailed in panels (C-E). 
Each average is taken over all time windows $\left[ t_0, t \right]$ shown in Fig. \ref{fig:mono_xcorr_full}, and averages are denoted by $\langle p \rangle_{t, t_0}$ for convenience.
(G) Structural response is asymmetric with respect to shear direction; this asymmetry is quantified by the ratio $P_p (2\omega^*)/P_p (\omega^*)$ for all shielding layers as a function of $\gamma_0$. 
Spectral densities $P_p(\omega)$ are calculated from $p(s, t \vert s, t-0.5)$. 
Each value is the mean of a set of $P_p(\omega)$ values calculated over consecutive $n$ cycle windows of $p(s, t \vert s, t-0.5)$ sliced through the full two-dimensional distribution shown in Fig. \ref{fig:mono_xcorr_full}.
We set $n = (5,7,7,7,7,2)$ for systems ordered by increasing strain amplitude.
Error bars for the ratio are calculated via Taylor series propagation of the standard deviations of the mean of each $P_p (\omega)$ quantity in the ratio.
Signals are colored by shielding level.
Note that for the most shielded layer, corresponding to the reddest color, $p(x, t \vert R_i, t - 0.5) = 1$ for all $t$ at some strain amplitudes. In this case, the ratio $P_p (2\omega^*)/P_p (\omega^*)$ is undefined, and thus not shown.
}
\label{fig:mono}
\end{figure*}

\begin{figure*}
\centering
\includegraphics[width=\textwidth]{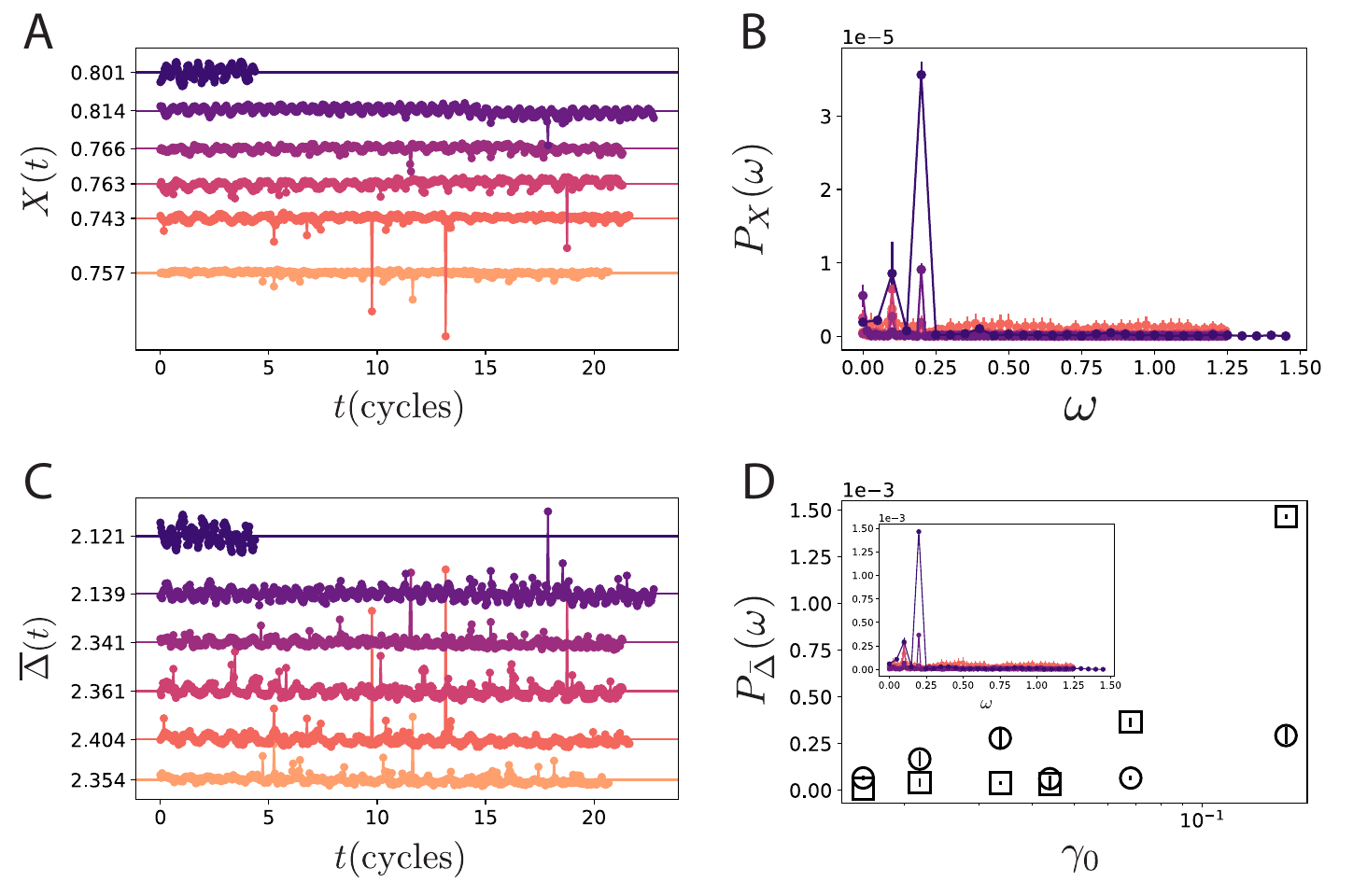}
\caption{
Global signatures of structural homogeneity for whole mono-disperse non-transient trajectories as a function of strain amplitude $\gamma_0$.
(A) Crystallinity in all systems. Signatures are arbitrarily offset for clarity; higher curves, with darker colors, correspond to higher values of $\gamma_0$. Horizontal lines indicate the mean of each crystallinity signature.
(B) Power spectral densities $P_X(\omega)$ of each crystallinity signature at all strain amplitudes.
Error bars are detailed in the caption of Fig. \ref{fig:mono}.
(C) The quantity $\overline{\Delta} (t)$ in $\mu m$, averaged over all particles in each system snapshot (not just those preserved during the whole trajectory). Signatures are arbitrarily offset for clarity; higher curves, with darker colors, correspond to higher values of $\gamma_0$. Horizontal lines indicate the mean of each signature.
(D) Two power spectral densities $P_{\bar{\Delta}}(\omega)$ of each $\overline{\Delta} (t)$ signature as a function of $\gamma_0$, for two distinct frequencies. Circles correspond to frequency $\omega^*$, the frequency of the needle oscillation, and squares correspond to frequency $2\omega^*$, or the second harmonic of the needle oscillation. 
Error bars are detailed in the caption of Fig. \ref{fig:mono}.
Inset are full power spectral densities $P_{\bar{\Delta}}(\omega)$ for all $\omega$ at all strain amplitudes.
}
\label{fig:mono_xtal_full}
\end{figure*}

\begin{figure*}
\centering
\includegraphics[width=\textwidth]{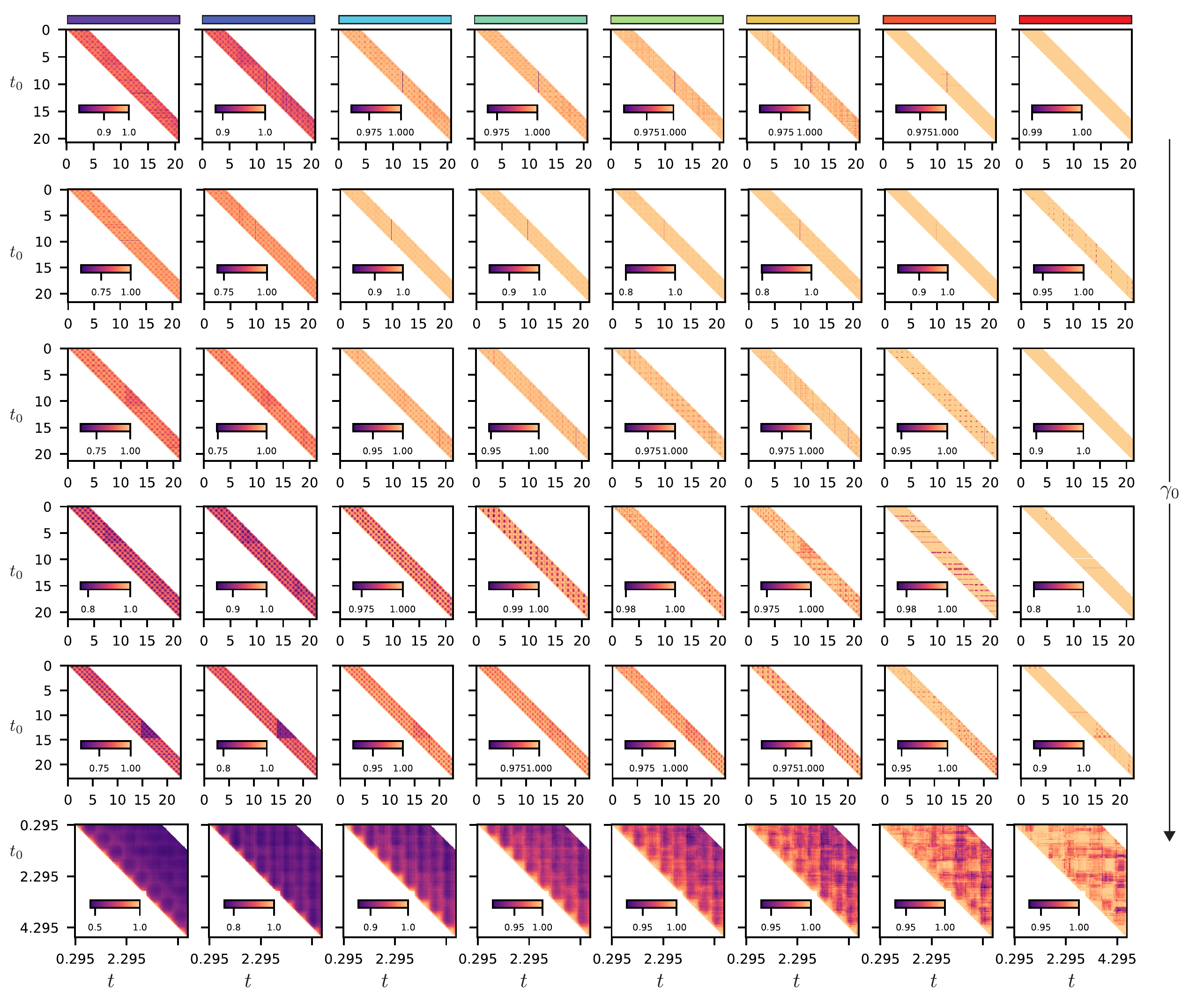}
\caption{Distributions of $p(s,t \vert s, t_0)$, with colored bars above each plot indicating the shielding level, for all mono-disperse systems.
Systems are ordered in increasing strain amplitude $\gamma_0$ as indicated by the arrow.
Signals are only shown for the non-transient portions of the experimental trajectories.
}
\label{fig:mono_xcorr_full}
\end{figure*}

\bibliographystyle{apsrev4-2}
\bibliography{dsb_mat,misc}